\documentclass[fleqn,usenatbib]{mnras}

\usepackage{newtxtext,newtxmath}
\usepackage{amsmath}

\usepackage[T1]{fontenc}

\DeclareRobustCommand{\VAN}[3]{#2}
\let\VANthebibliography\thebibliography
\def\thebibliography{\DeclareRobustCommand{\VAN}[3]{##3}\VANthebibliography}

\usepackage{graphicx}	
\usepackage{amsmath}	

\title[X-ray variability of QSOs]{Ensemble X-ray variability of optically selected QSOs: dependence on black hole mass and Eddington ratio}

\author[A. Georgakakis et al.]{
A. Georgakakis,$^{1}$\thanks{E-mail: age@noa.gr}, 
J. Buchner$^{2, 3}$, 
A. Ruiz$^{1}$, 
T. Boller$^2$, 
A. Akylas$^{1}$, 
M. Paolillo$^{4, 5, 6}$, 
M. Salvato$^2$,\newauthor
A. Merloni$^2$,
K. Nandra$^2$,
T. Dwelly$^2$
\\
$^1$Institute for Astronomy \& Astrophysics, National Observatory of Athens, V.  Paulou  \& I.  Metaxa, 11532,  Greece\\
$^2$ Max Planck Institute for Extraterrestrial Physics, Giessenbachstrasse, 85741 Garching, Germany\\
$^3$Excellence Cluster Universe, Boltzmannstr. 2, D-85748, Garching, Germany\\
$^4$Dipartimento di Fisica "Ettore Pancini", Universit\`a Federico II, via Cinthia, 80126 ,Napoli, Italy\\
$^5$INAF - Osservatorio Astronomico di Capodimonte, via Moiariello 16, 80131, Napoli, Italy\\
$^6$Istituto Nazionale di Fisica Nucleare, Sezione di Napoli, I-80126 Napoli, Italy
}

\date{Accepted XXX. Received YYY; in original form ZZZ}

\pubyear{2024}

\begin{document}
\label{firstpage}
\pagerange{\pageref{firstpage}--\pageref{lastpage}}
\maketitle

\begin{abstract}
Although flux variability is one of the defining properties of accretion flows  onto supermassive black holes, its dependence on physical parameters such as the mass of the compact object and the Eddington ratio remain under discussion. In this paper we address this issue using the structure function statistic to measure the variability at X-ray wavelengths of a sample of optically selected QSOs with available black hole masses and Eddington ratios. We present a new Bayesian methodology for estimating the structure function tailored to the Poisson nature of the X-ray data. This is applied to 15,548 SDSS DRQ16 QSOs with repeat observations in the {\it XMM-Newton} archive and/or the SRG/eROSITA All Sky Survey. The X-ray structure function monotonically increases to time intervals of about 10-15\,yrs, consistent with scenarios in which instabilities of the accretion disk contribute to the X-ray variability on long timescales. Additionally, there is evidence that the amplitude of the stochastic X-ray flux variations rises with decreasing black hole mass and Eddington ratio. This finding imposes stringent constraints on empirical models of Active Galactic Nuclei variability derived from local samples, emphasizing the significance of high-redshift population studies for comprehending the stochastic flux variations in active black holes.
\end{abstract}




\section{Introduction}

The accretion of material onto supermassive black holes (SMBHs) at the centres of galaxies is characterised by a high level of stochasticity, which imprints temporal flux variations of the emitted radiation over a wide range of time scales. This fundamental property of the accretion process provides an important observational tool to explore the innermost regions of active SMBHs at spatial scales that are challenging to resolve by other means. This is because flux variations can be translated to light travel times and hence, provide information on the dimensions, spatial structure and dynamics of the accretion flow onto SMBHs \citep[e.g.][]{Uttley2014, Cackett2021}. This potential has led to systematic multiwavelength studies to better understand the statistical properties and nature of the stochastic flux variations of accretion events onto the SMBHs, i.e. Active Galactic Nuclei (AGN).  A central theme in these studies is the dependence of the variability on the fundamental physical parameters of the accreting system, e.g. black hole mass and Eddington ratio. Scaling relations between these physical quantities and the statistical properties of the observed temporal variations would allow the study of accreting systems over a wide range of black hole masses and accretion rates under the same theoretical paradigm. 

A popular method to explore the statistical properties of the stochastic AGN flux variability is the Power Spectral Density (PSD), which describes the distribution of the variance of a light curve in Fourier frequencies.  X-ray monitoring campaigns of nearby Seyferts over long periods of time reveal PSDs that can be typically described by a broken power-law with slopes changing from about --2 above a break frequency to  about --1 below it, similar to X-ray stellar binaries \citep[e.g.][]{Uttley2002, Papadakis2002, Papadakis2003, Markowitz2003, McHardy2004, Gonzalez-Martin_Vaughan2012}. The position of the PSD break frequency is shown to scale with both the mass of the black hole and the Eddington ratio. More importantly, this trend can also be extended to X-ray stellar binaries with black holes many orders of magnitude less massive than AGN \citep[e.g.][]{McHardy2006}. Unfortunately, the direct estimation of the PSD is limited to a small number of nearby bright AGN, for which long and uninterrupted campaigns, mostly from the  Rossi X-ray Timing Explorer (RXTE), are available. One approach to overcome this limitation and enable the statistical characterisation of the variability of large samples of AGN is by estimating the normalised excess variance of light curves \citep{Nandra1997}. This quantity is mathematically related to the integral of the PSD over the time-scales sampled by a light curve. It can be measured even in the case of low signal-to-noise observations or not well-sampled light curves \citep[e.g.][]{Allevato2013}. The rich archives of the {\it XMM-Newton} and {\it NuSTAR} X-ray telescopes has enabled the determination of the normalised excess variance at different timescales for large samples of low redshift AGN   \cite[e.g.][]{Ponti2012, Akylas2022, Tortosa2023}. These studies reveal correlations of the excess variance, and hence the PSD, on black-hole mass and possibly Eddington ratio, in broad agreement with earlier results based on the PSD analysis of a few local AGN. 

Parallel to studies of the light curves of individual objects, there have been efforts to characterise the mean variability properties of high redshift AGN populations detected in extragalactic X-ray survey fields to constrain their PSDs \citep{Paolillo2004, Papadakis2008, Paolillo2017, Paolillo2023}. This approach hinges on the fact that the total integration time of at least some X-ray surveys has been gradually built up by numerous repeat observations carried out over the course of many years. The light curves of individual sources in such surveys carry limited information but when averaged together they provide useful constraints on the PSD of the population. This approach shows that distant AGN have variability properties that scale with black hole mass and Eddington ratio in a similar fashion as their local counterparts \citep{Georgakakis2021, Paolillo2023}. However, the exact dependence of the variability on the physical parameters of the accreting systems is under investigation. The \cite{Georgakakis2021} modeling favors a bending power-law PSD with a break frequency that scale with black hole mass \citep{Papadakis2004} and an amplitude that anti-correlates with Eddington ratio \citep{Ponti2012}. The \cite{Paolillo2023} analysis is consistent with a bent power-law PSD model with constant amplitude and a brake frequency that depends on black hole mass and possibly Eddington ratio \cite{McHardy2006}.

In this paper we adopt the ensemble variability approach to further explore  the PSD of distant optically selected QSOs and get a handle on scaling relations with black hole mass and Eddington ratio. The variability statistic we chose to use is the structure function \citep{Kozlowski2016, Kozlowski2017}. It provides a measure of the flux variance at a given time scale based on two-epoch photometric measurements for a large sample of sources \citep[e.g.][]{deVries2005, MacLeod2012A, Vagnetti2016, Middei2017}. We combine the SDSS DRQ16 QSO sample with {\it XMM-Newton} archival data and new observations from the eROSITA \citep[extended ROentgen Survey with an Imaging Telescope Array,][]{Predehl2021} instrument onboard the SRG (Spectrum-Roentgen-Gamma) spacecraft to measure the X-ray structure function to timescales of about 10\,years at the rest-frame of the sources. Our analysis is similar to previous studies by \cite{Vagnetti2016} and  \cite{Middei2017} but our larger sample size enables grouping sources by black-hole mass and Eddington ratio to investigate systematic variations of the structure function with these parameters. Moreover, we also present a new Bayesian algorithm for the estimation the structure function that robustly accounts for X-ray upper limits by modeling the Poisson nature of the X-ray data. Finally the interpretation of our results is supported by the light curve simulation tools presented by \cite{Sartori2018, Sartori2019}.

The analysis presented in this work is similar to the recently accepted paper by \cite{Prokhorenko2024}. They also explore the structure function of SDSS QSOs by combining {\it XMM-Newton} observations with the eROSITA data in the part of the sky with Galactic longitudes $0 < l < 180$\,deg, in which the proprietary rights lie with the Russian SRG/eROSITA team. There are however, a number of differences between our work and that  of \cite{Prokhorenko2024}. Our QSO sample is selected from the latest SDSS DRQ16 catalogue \citep[][see Section \ref{sec:sample}]{Lyke2020} instead of the  DRQ14 one \citep{Paris2018} used by \cite{Prokhorenko2024}. The structure function calculation in our study is based on 2-epoch X-ray observations of SDSS QSOs, in which at least one epoch is from {\it XMM-Newton} and the second is either from {\it XMM-Newton} (i.e. repeat observations) or eROSITA. Instead \cite{Prokhorenko2024} use only  {\it XMM-Newton}/eROSITA pairs in their calculations. As a result of the points above the sample used here numbers approximately 15,500 unique QSOS (see Section \ref{sec:sample}) as opposed to 2344 used by \cite{Prokhorenko2024}. We also re-iterate that our analysis uses on a new Bayesian methodology for calculating the structure function (see Section \ref{sec:SF}) that is adapted to the Poisson nature of the X-ray observations. 

Throughout this work, we adopt a flux $\Lambda$-CDM (Cold Dark Matter) cosmology with $\rm H_{0} = 70 km \, s^{-1}$  and $\Omega_\Lambda= 0.7$.

\section{The Sample}\label{sec:sample} 

The AGN sample used in this work is selected from the Sloan Digital Sky Survey (SDSS) quasar catalogue data release 16 \citep[DR16Q][]{Lyke2020}. It provides photometric and spectroscopic information for 750,414 spectroscopically confirmed QSOs targeted by the SDSS. Black hole masses, bolometric luminosities and Eddington ratios for the DRQ16 QSOs are taken from \cite{Wu_Shen2022}. 
 
In this work we explore the X-ray variability properties of the DR16Q QSOs via their structure function (see Section \ref{sec:SF}). For this calculation X-ray flux estimates for  individual  sources  at  two  distinct  epochs  are  required.  Two independent datasets are used for the determination of the structure function. The first is based on repeat observations in the {\it XMM-Newton} archive. The second is combining {\it XMM-Newton} data with the German eROSITA All Sky Survey Data Release 1 \citep[eRASS-DE DR1;][]{Merloni2024}. The latter dataset consists of observations carried out in the first six months of the SRG/eROSITA all-sky survey (eRASS1) whose proprietary rights lie with the German eROSITA consortium (eROSITA-DE).  The statistical methodology adopted in this work to infer the X-ray structure function of DR16Q QSOs  does not rely only on X-ray detected sources either in the {\it XMM-Newton} archive or the eRASS-DE DR1. Instead we perform aperture X-ray photometry at the positions of  all DR16Q QSOs and then feed this information into the Bayesian model described in Section \ref{sec:SF} to determine the structure function of the population at fixed rest-frame timescales. In the case of the {\it XMM-Newton} the RapidXMM database \citep{Ruiz2022} is used to query aperture photometry products in the 0.2-2\,keV spectral band (X-ray photons, background level, mean exposure time) at the positions of  DR16Q QSOs. The RapidXMM aperture size is fixed to 15\,arcsec radius in the case of pointed {\it XMM-Newton} observations. This radius roughly corresponds to about 70\% of the Encircled Energy Fraction (EEF) of the {\it XMM-Newton} PSF \citep[][]{Ruiz2022}. For eRASS-DE DR1 the \texttt{apetool} task of the eROSITA Science Analysis Software System \citep[eSASS,][]{Brunner2022} is used to extract X-ray photon counts, estimate the background level and corresponding mean exposure time in the 0.2-2.3\,keV energy band at the positions of DR16Q QSOs that overlap with the eRASS-DE DR1 footprint. The adopted aperture size in this case corresponds to the 75\% EEF of the eROSITA PSF at the position of interest. 

There are a total of 12,337 unique SDSS QSOs with at least two independent observations in the {\it XMM-Newton} archive. This number excludes sources (total of 2116) that  lie close to either CCD gaps of the EPIC \cite[European Photon Imaging Camera;][]{Turner2001} detectors or the edge of the field of view of the {\it XMM-Newton}. The identification of such sources is based on the RapidXMM flagging scheme \citep[see Table 2 of ][flagbits 0 and 1]{Ruiz2022}. Contamination of the RapidXMM photometric products by photons associated with the  wings of the PSF of nearby X-ray sources is an issue that can potentially bias the analysis. This is addressed by first cross-matching the optical positions of the 12,337  SDSS QSOs with the 4XMM-DR12 serendipitous X-ray source catalogue \citep{Webb2020} within a radius of 6\,arcsec. This yield a total of 9474 associations. For the density of the 4XMM-DR12 X-ray detections (630,347 unique sources over an area of $\rm 1283\,deg^2$) and the adopted matching radius the expected chance association rate is 0.4\%. We then search for additional QSOs in the DRQ16 catalogue that lie close to (within 45\,arcsec) 4XMM-DR12 X-ray sources but are not associated with them (i.e. they do not belong to the subsample of the 9474 4XMM-DR13/DRQ16 identifications).  The X-ray aperture photometry of such QSOs may be contaminated by photons associated with the PSF wings of the nearby X-ray detections. There are 598 DRQ16 QSOs that lie within 45\,arcsec off the positions of 4XMM-DR12 X-ray sources. These are removed from the analysis. This results to a sample of 11,739 SDSS QSOs with a least two independent {\it XMM-Newton} observations.

The overlap between eRASS-DE DR1 footprint and the {\it XMM-Newon} archive yields  a total of 9075 unique  DRQ16 QSOs with  at least one pair of observations in the two datasets. This number excludes systems (total of 430) located close to CCD gaps of {\it XMM-Newton} selected as explained above. We further identify SDSS QSOs that are in the vicinity to X-ray detections in either the {\it XMM-Newton} data or the eRASS-DE DR1 catalogue but are not associated with them. In the case of the {\it XMM-Newton}, such sources are selected as explained above.  In the case of the eROSITA data, we define exclusion regions of 45\,arcsec around all eRASS1 X-ray sources with detection likelihood  {\sc det\_ml}$>7$ and reject any SDSS QSOs that overlap with them. In this exercise we also use the eRASS1 optical identification catalogue presented by Salvato et al. (2024 in prep.) to identify a total of 896 unique DR16 SDSS QSOs that are associated with eRASS-DE DR1 X-ray detections in the Legacy Survey \citep{Dey2019} Data Release 10 footprint. These sources are allowed in the structure function calculation. The selections above result to 8499 unqiue SDSS QSOs with at least two independent {\it XMM-Newton} and eRASS-DE DR1 observations. The total number of unique DRQ16 QSOs with at least two independent {\it XMM-Newton} observations or independent {\it XMM-Newton} and eRASS-DE DR1 observations is 17,940. 

X-ray diffuse emission associated with the hot gas of clusters of galaxies may contaminate the aperture photometry at the positions of SDSS QSOs, particularly in the case of the broad PSF of eROSITA ($\approx 30$\,arcsec HEW). We therefore remove SDSS QSOs that lie in the close vicinity of known X-ray selected clusters (see Section 5.2 of \citealt{Merloni2024}) culled from the ROSAT All-sky Survey \citep[RXGCC,][]{Xu2022}, the XMM Cluster Survey \citep[XCS,][]{Mehrtens2012},  the XMM CLuster Archive Super Survey \citep[X-CLASS,][]{Clerc2012}, the XMM-XXL survey \citep[XXL365,][]{Adami2018}, the eROSITA Final depth Equatorial Survey \cite[eFEDS,][]{Liu2022}. Each of these clusters is assigned an exclusion radius that is empirically determined and is different for the {\it XMM-Newton} and eROSITA observations because of the different PSF of the two telescopes. The choice of the size of the exclusion radius is a trade-off between large QSO sample and low level of contamination from X-ray photons associated with cluster emission. For the eROSITA observations the exclusion radius is defined as $0.5 \times R_{500}$  in the case of systems with measured $R_{500}$ parameter, or 500\,kpc in the case of clusters with no $R_{500}$ estimates, or 1\,arcmin for systems with unknown redshifts. For the {\it XMM-Newton} observations the exclusion radius is defined as $0.15 \times R_{500}$  in the case of systems with measured $R_{500}$ parameter, or 150\,kpc in the case of clusters with no $R_{500}$ estimates, or 0.5\,arcmin for systems with unknown redshifts. This step filters out a total of 1573 unique SDSS QSOs. 

SDSS QSOs associated with blazars may show X-ray variability that is not stochastic. We therefore also remove 79 SDSS QSOs that lie within 6\,arcsec off blazars taken from the 5th edition of the Roma-BZCAT catalogue \citep{Massaro2015}. Blazars are a subset of the class of radio loud QSOs, the X-ray flux of which is often dominated by emission from the radio jets. As a result the  high energy temporal properties of such sources may not be directly related to the  X-ray corona and the accretion flow onto the central supermassive black hole. We quantify the radio loudness of a source via the relation \citep{Kellermann1989}

\begin{equation}
    R_F =\frac{F_{\rm 5\,GHz}}{F_{\rm 4400\,\mathring{A}}}
\end{equation}

\noindent where $F_{\rm 5\,GHz}$ and $F_{\rm 4400\,\mathring{A}}$ are the monochromatic flux densities at 5\,GHz and 4400\,\AA\, respectively. For the determination of the $F_{\rm 4400\,\mathring{A}}$ we first estimate the flux density at rest-frame $\rm 2500\,\mathring{A}$ using the apparent $i$-band magnitude as explained in Section 5 of \cite{Richards2006}. The flux density at rest-frame $\rm 2500\,\mathring{A}$ is converted to $F_{\rm 4400\,\mathring{A}}$ assuming  a power-law spectral energy distribution of the form 
$F_\nu \propto \nu^{-0.44}$. For the estimation of $F_{\rm 5\,GHz}$ we use two different large-area radio surveys. The  FIRST \citep[Faint Images of the Radio Sky at Twenty centimeters][]{Becker1995} survey and the 2nd data release of LoTSS \citep[LOFAR Two-metre Sky Survey,][]{Shimwell2017, Shimwell2022}.  The former covers about 10,000\,deg$^2$ and most of the SDSS area to a median 1$\sigma$  rms sensitivity of  0.141$\rm \,mJy\,beam^{-1}$ at 1.4\,GHz. The LoTSS DR2 currently covers about 5,700\,deg$^2$ at 144\,MHz to a deeper flux density limit compared to FIRST with a median 1$\sigma$  rms sensitivity of 83\,$\rm \mu Jy \, beam^{-1}$. The  SDSS DRQ16 catalogue has already been matched to  the positions of sources detected in FIRST. 
For QSOs without FIRST counterparts we estimate $3\sigma$ upper limits to their 1.4\,GHz flux density using the FIRST RMS noise maps\footnote{sundog.stsci.edu/first/catalogs/readme.html\#coverage}. These upper limits are given by the   relation $0.25 + 3\sigma_{RMS}$, where $\sigma_{RMS}$ is the FIRST RMS noise at the source position and 0.25 is the CLEAN bias correction \citep{White1997}. The association of the SDSS DRQ16 QSOs with the LoTSS catalogues uses the sky positions of the optical counterparts of LoTSS sources presented by  \cite{Hardcastle2023}. For QSOs without LoTSS matches we adopt an approximate $3\sigma$ upper limit to their 144\,MHz flux densities of 0.3\,mJy based on the median RMS noise of 83\,$\rm \mu Jy \, beam^{-1}$. For both radio surveys a power-law radio spectral shape with index $-0.8$ \citep[i.e. $F_\nu \propto \nu^{-0.8}$;][]{Sanchez-Saez2018} is assumed to convert the observed FIRST 1.4\,GHz and/or LoTSS 144\,MHz flux densities to rest-frame ones at 5\,GHz. 

\begin{figure}
	\includegraphics[width=\columnwidth]{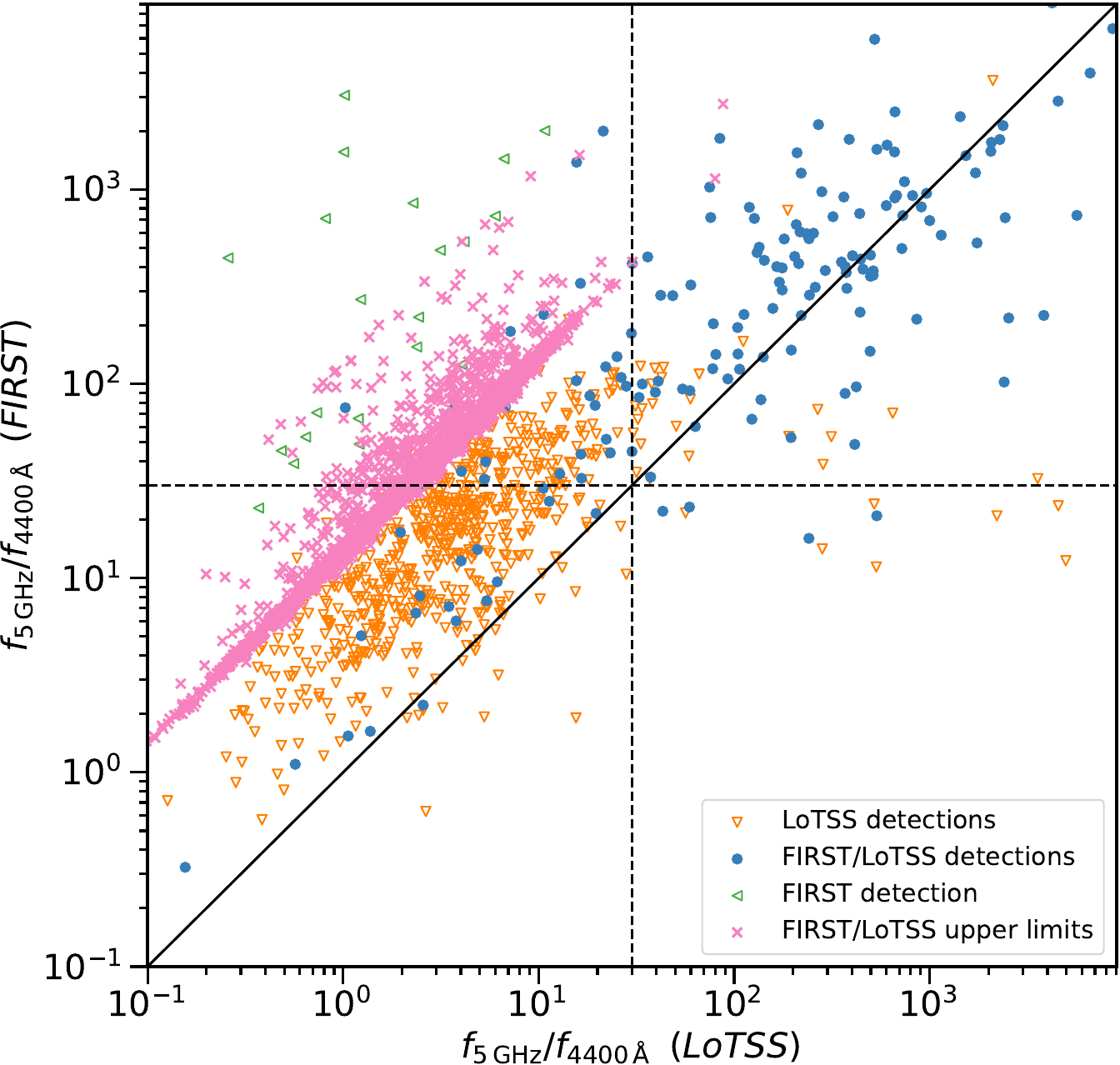}
    \caption{Comparison of the radio loudness parameter (see Section \ref{sec:sample}) estimated using either the FIRST data (y-axis) or the LOFAR LoTSS survey (x-axis). All the data points are SDSS DR16 QSOs that lie in the overlap region of two radio samples. The black diagonal solid line shows the one-to-one relation. The horizontal and vertical dashed lines show the radio loudness threshold $R_F=30$ adopted in our work to separate radio loud from radio quiet sources.  Filled blue circles represent QSOs detected in both radio surveys. The orange open triangles pointing downward are SDSS QSOs that are detected in LoTSS but not in FIRST. The green open triangles pointing to the left correspond to SDSS QSOs that are detected in FIRST but not in LoTSS. The pink crosses are SDSS QSOs with upper limit radio flux density measurements in both surveys. They are offset from the one-to-one relation because of the different depths of the FIRST and LoTSS radio surveys.}
    \label{fig:RLvsLR}
\end{figure}

The traditionally adopted cut for defining radio quiet AGN is $R_F<10$ \citep{Kellermann1989}.  Recent observational evidence nevertheless suggests that the contribution of the radio jet emission to the X-ray flux of AGN is subdominant until $R_F\ga100$ \citep[e.g][]{Zhu2020}. In this work we therefore adopt the less conservative threshold of  $R_F=30$ to differentiate radio loud from radio quiet DRQ16 QSOs \citep{Timlin2020}. Figure \ref{fig:RLvsLR} compares the radio loudness parameters of DRQ16 QSOs estimated using the FIRST and LoTSS flux densities. This figure demonstrates that the LoTSS radio loudness upper limits provide much stronger constraints on the radio properties of SDSS QSOs compared to the shallower FIRST survey. For example, there are 6083 unique DRQ16 QSOs that have both FIRST and LoTSS coverage. From this sub-sample a total of 3410  have FIRST radio loudness upper limits that lie above our adopted threshold $R_L=30$. For these sources it is not possible to conclude, based on the FIRST data alone, if they are radio quiet. The deeper LoTSS data however, yield $R_L$ upper limits or measurements that are below the $R_L=30$ threshold for 98\% of these sources (3350 out of the 3410). These numbers show that the DRQ16 sub-sample without FIRST associations is dominated by radio quiet QSOs. We therefore define as radio loud those DRQ16 sources that are associated with radio detections in either FIRST or LoTSS and have $R_L>30$ in either of these surveys. These sources are exclude from the analysis. Our sample may contain residual radio loud QSOs associated with FIRST non-detections. Such sources however. represent less than 2\% of the total sample and are unlikely to have an impact on our results and conclusions. 

It is recognised that the redshift estimates of the SDSS DRQ16 QSO catalogue become uncertain toward fainter optical magnitudes and lower signal-to-noise optical spectra \citep[e.g.][]{Menzel2016}. The DRQ16 catalogue {\sc zwarning} flag produced by the SDSS spectral reduction pipeline and the {\sc sn\_median\_all} parameter, which provides an estimate of the signal-to-noise ratio of  individual spectra, can be used to identify potentially problematic sources. A conservative approach proposed by \cite{Menzel2016} to select highly reliable redshifts is to retain only sources with {\sc zwarning=0} and {\sc sn\_median\_all$>1.6$}. This approach removes about 30\% of the sample, inlcuding a large fraction of sources with bona-fide and reliable QSO redshifts (based on visual inspection of a subsample). In this work we therefore choose not to apply a redshift quality selection. Instead, we repeat the analysis presented in the following section for the sample with {\sc zwarning=0} and {\sc sn\_median\_all$>1.6$} and confirm that the main trends and results of Section \ref{sec:results} are recovered for that conservative sample.

\begin{figure}
	\includegraphics[width=\columnwidth]{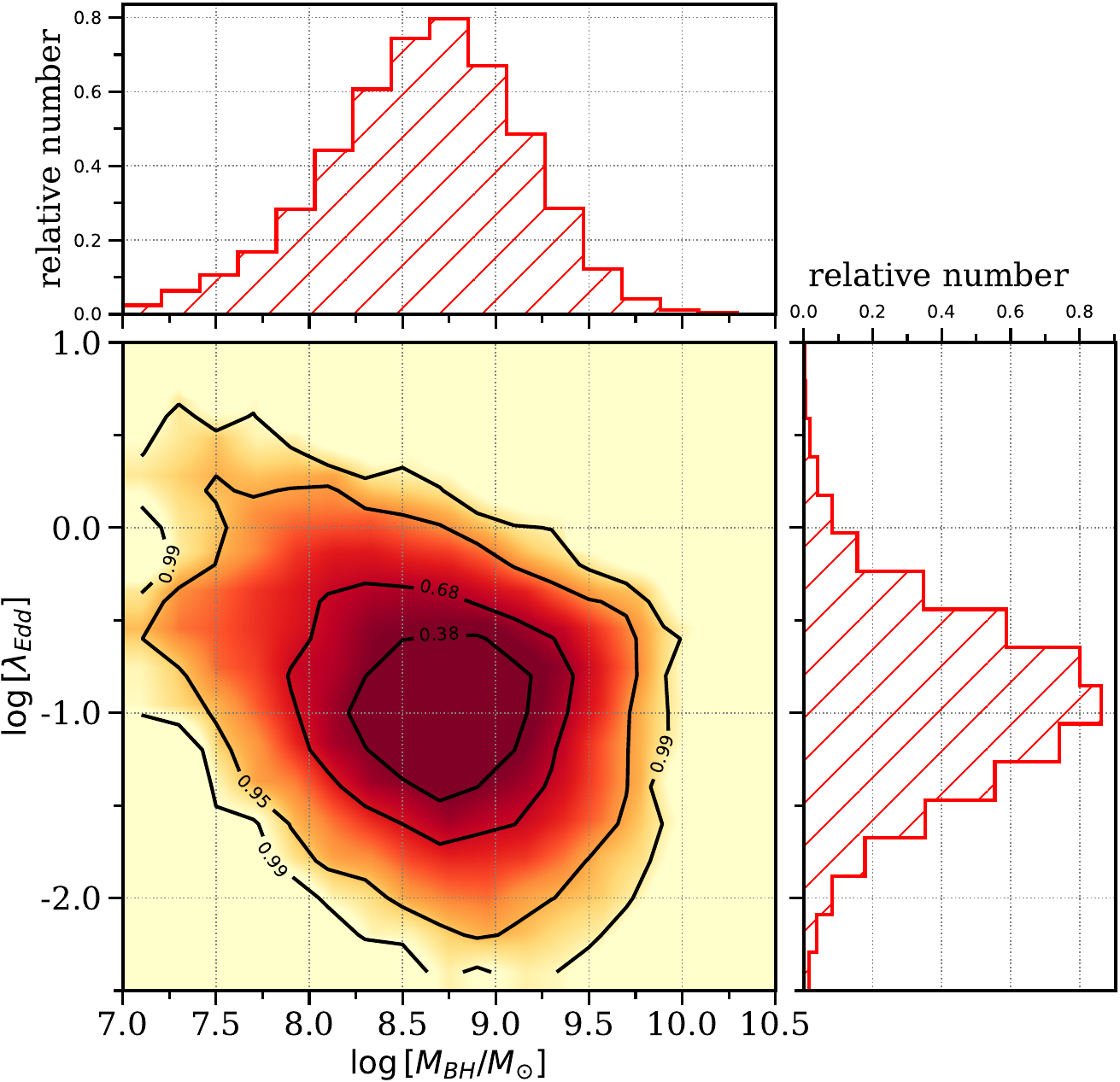}
    \caption{Distribution of the baseline sample of SDSS DRQ16 QSOs (see Section \ref{sec:sample}) on the Eddington ratio vs black hole mass plane. The contour levels are chosen to enclose 34, 68, 95 and 99 per cent of the QSO population. The 1-dimensional projections of this distribution along the black hole mass and Eddington ratio axes are also shown by the histograms on the top and to the right of the main panel respectively.}
    \label{fig:MBHLEDD-dist}
\end{figure}

\begin{figure}
	\includegraphics[width=\columnwidth]{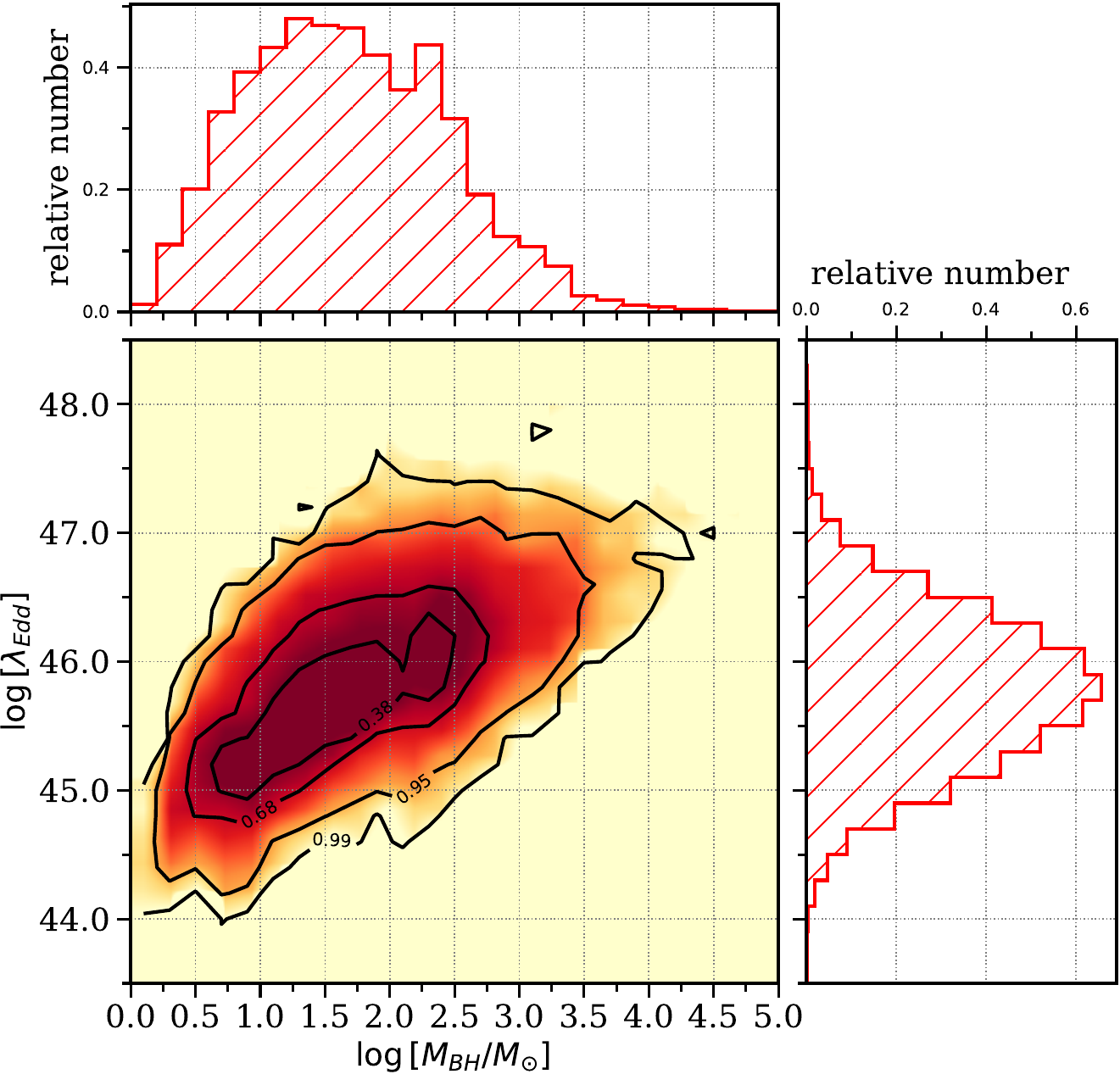}
    \caption{Distribution of the baseline sample of SDSS DRQ16 QSOs (see Section \ref{sec:sample}) on the bolometric luminosity vs redshift plane. The contour levels are chosen to enclose 34, 68, 95 and 99 per cent of the QSO population.
    The 1-dimensional projections of this distribution along the bolometric luminosity and redshift axes are also shown by the histograms on the right and at the top of the main panel respectively.}
    \label{fig:LBOLZ-dist}
\end{figure}

\begin{figure}
	\includegraphics[width=\columnwidth]{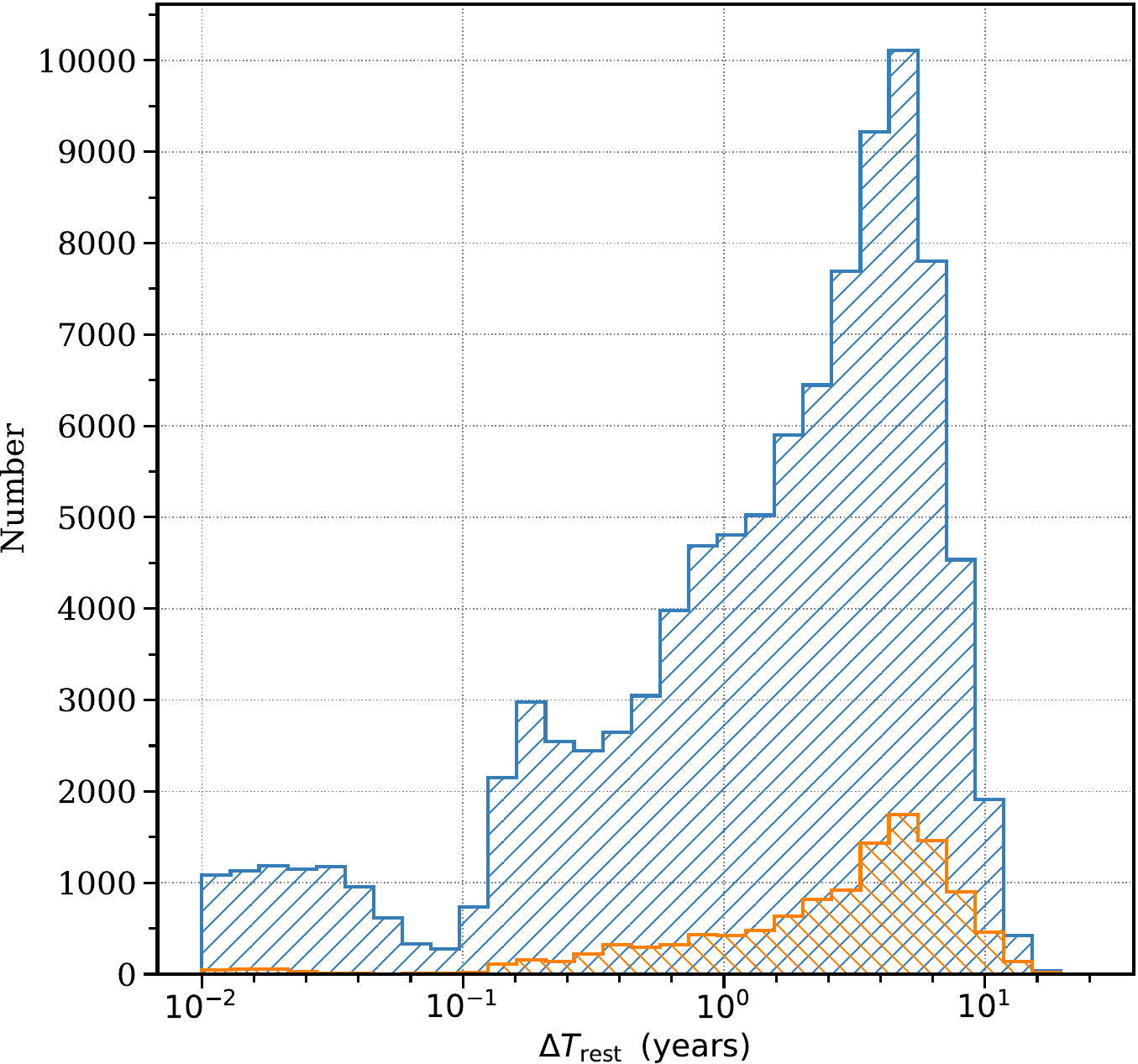}
    \caption{Histogram of the rest-frame time scales probed by the baseline sample of SDSS DRQ16 QSOs (see Section \ref{sec:sample}). These timescales are estimated as the time difference between either {\it XMM-Newton} repeat observations or {\it XMM-Newton} and the eRASS-DE DR1 data at the rest-frame of each QSO. The blue hatched histogram corresponds to the full sample. The orange histogram is for the subsample that combines {\it XMM-Newton} observations with the eRASS-DE DR1 data. The contribution of the latter sample to the total increases with increasing timescale.}
    \label{fig:DT-dist}
\end{figure}

Following the selection criteria described above, the final (baseline) sample used in this paper to determine the X-ray structure function of SDSS DRQ16 QSOs consists of a total of 15,548 unique QSOs  and 110,829 pairs of X-ray observations at distinct epochs. Figure \ref{fig:MBHLEDD-dist} shows the distribution of the baseline subsample  on the Eddington ratio vs black hole mass plane. In Figure \ref{fig:LBOLZ-dist} we present the distribution of the sample on the bolometric luminosity vs redshift plane. Figure \ref{fig:DT-dist} shows the histogram of the rest-frame time differences, $\Delta T$, of the sample. This quantity is defined as the time span between the two epoch X-ray observations of a given source divided by the factor $(1+z)$, where $z$ is the redshift of the source.  Figure \ref{fig:DT-dist} shows that the sample is probing timescales that extend to about 1\,decade at rest-frame. Also, the contribution of the XMM/eROSITA flux measurement pairs increases toward longer timescales. In our analysis we use rest-frame timescales above the limit $10^{-2}$\,yr (i.e. few days).  For shorter time intervals the expected flux differences between the two-epoch observations are small and systematics (e.g. uncertainties in the estimation of the encircled energy fraction for the RapidXMM counts) become important. In any case Figure \ref{fig:DT-dist} shows that the statistical power of the sample is for time intervals ranging from months to several years. These are the timescales of interest to this work.

\section{Structure Function Estimation}\label{sec:SF}

Multiple definitions of the structure function (SF) exist in the literature depending on whether the data are in flux or apparent magnitude units \citep{Kozlowski2017}. In this work we define the structure function at a given time scale $\tau$ as 

\begin{equation}\label{eq:sf-basic}
SF^2(\tau)=\langle [ \log f_{X}(t) - \log f_{X}(t+\tau) ]^{2}\rangle, 
\end{equation}

\noindent where $f_{X}(t)$, $f_{X}(t+\tau)$ are the X-ray fluxes of a  source measured at epochs separated by $\tau$, i.e. at times $t$ and $t+\tau$. For a population of AGN the equation above estimates the variance of the logarithmic flux differences at the timescale $\tau$. In practice observations are noisy and therefore any contribution from random flux errors must be accounted for in the SF estimation. This is usually done by modifying Equation \ref{eq:sf-basic} as \citep[e.g. see][]{Middei2017, Kozlowski2017}

\begin{equation}
SF^2(\tau)= \langle [ \log f_{X}(t) - \log f_{X}(t+\tau) ]^{2}\rangle - \langle \sigma^2_X(t) + \sigma^2_{X}(t+\tau)\rangle,
\end{equation}
 
\noindent where $\sigma^2_X(t)$ represents the uncertainty associated with the flux measurement $\log f_{X}(t)$. X-ray observations are typically described by the Poisson distribution and therefore the  shot noise term above depends on the instantaneous source flux and the instrumental background level. In the case of bright sources the errors may be approximated to an acceptable level of accuracy by the normal distribution. In the case of low photon counts however, such as the eRASS1 observations or shallow {\it XMM-Newton} data, this approximation breaks down. For example, many of the SDSS DRQ16 QSOs are not formally detected in the eRASS1 (see Section \ref{sec:sample}) resulting in upper flux limits, which may nevertheless  provide useful constraints on the SF. For the calculation of the SF we therefore develop a custom Bayesian methodology that properly accounts for the Poisson nature of the data and allows the consistent treatment of both formal X-ray detections and X-ray upper limits. At the core of the methodology is aperture photometry as described below. 

Suppose a source $i$ with X-ray observations in two distinct epochs.  In the first dataset the source has $N_{i,1}$ counts within an aperture with EEF $EEF_{i, 1}$. The background level is assumed to be $B_{i,1}$ and the exposure time at the source position is $t_{i,1}$. The probability that the source has flux $f_{i,1}$ can be expressed by the Poisson probability

\begin{equation}\label{eq:prob-epoch1}
    P(f_{i,1} \,|\, O_{1}) = \frac{e^{-\lambda_{i,1}}\times\,\lambda_{i,1}^{N_{i,1}}}{N_{i,1}!}.
\end{equation}

\noindent  Where $O_1$ signifies the observations at hand (e.g. $N_{i,1}$, $B_{i,1}$ etc) and the Poisson expectation value $\lambda_{i,1}$ is
\begin{equation}
    \lambda_{i,1}=f_{i,1}\times ECF_{i,1}\times t_{i,1}\times EEF_{i,1}\,+B_{i,1}.
\end{equation}

\noindent $ECF_{i, 1}$ is the energy to flux conversion factor and depends on the source's spectral model and the characteristics of the X-ray telescope/detector that observed it. In the second X-ray observation carried out at a later epoch by the same or different telescope the source $i$ has $N_{i, 2}$ counts within an aperture of EEF $EEF_{i, 2}$, the background level is $B_{i, 2}$ and the exposure time is $t_{i,2}$. The probability that the source’s flux has changed by $\Delta \log F_i  = \Delta \log (f_{i, 1}/f_{i, 2}) = \log f_{i,1} - \log f_{i,2}$ to $f_{i,2}$ is also given by the Poisson formula

\begin{equation}\label{eq:prob-epoch2}
    P(f_{i,1}\cdot\frac{f_{i,2}\,}{f_{i,1}} \,|\, O_{2} )\,
    =\,\frac{e^{-\lambda_{i,2}}\cdot\,\lambda_{i,2}^{N_{i,2}}}{N_{i,2}!}.
\end{equation}

\noindent  Here $O_2$  signifies the second set of observations at hand. The flux variation is expressed by the ratio $f_{i,1}/f_{i,2}$, which links to the definition of the structure function in Equation \ref{eq:sf-basic}. The Poisson expectation value $\lambda_{i,2}$ is

\begin{equation}
    \lambda_{i,2}=\left(f_{i,1}\cdot \frac{f_{i,2}}{f_{i,1}}\right)\times ECF_{i,2}\times t_{i,2}\times EEF_{i,2}\,+B_{i,2}.
\end{equation}

\noindent Where $ECF_{i, 1}$ is the energy to flux conversion factor for the second observation. The likelihood of the two epoch observations of source $i$ is the product of the Poisson probabilities of Equations \ref{eq:prob-epoch1}, \ref{eq:prob-epoch2} and the probability of a flux ratio variation $f_{i,2}/f_{i,1}$. For the latter, it is assumed that the variations of the logarithmic flux ratio, $\log f_{i,2}/f_{i,1}$, are drawn from a normal distribution with mean of zero and variance that depends on the (rest-frame) time interval $\Delta T$ between the two observations. As a consequence of Equation \ref{eq:sf-basic}, the variance of the adopted normal distribution is by definition the structure function at timescale $\Delta T$, which is to be constrained from the observations. Combining all the components above it is possible to write down the likelihood for the 2-epoch observations of source $i$ as

\begin{equation}
\mathcal{L}_{i} = P(f_{i,1} \,|\, O_{1} ) \cdot P( f_{i,1}\cdot\frac{f_{i.2}}{f_{i,1}} \, | \, O_{2} ) \cdot \mathcal{N}\left( \log \frac{f_{i.2}}{f_{i,1}} \,| \, \mu,\sigma \right)
\end{equation}

\noindent The last term describes the normal distribution with parameters $\mu$ (mean) and $\sigma$ (scatter). In the case of $N$ number of sources with two epoch observations separated by the same time interval $\Delta T$ the likelihood of the population is

\begin{equation}\label{eq:likelihood}
\begin{split}
    \mathcal{L}= \prod_{i=1}^{N}\mathcal{L}_{i} =  
                 \prod_{i=1}^{N} & P\left( \log f_{i,1} \, | \, O_{1}\right) \,\cdot \\
                 & P[ \log f_{i,1} + \log (f_{i,2}/f_{i,1}) \, |\, O_{2}] \cdot\\
                 &\mathcal{N}[\log (f_{i,2}/f_{i,1}) \,| \, \mu,\sigma ]
\end{split}
\end{equation}

\noindent The likelihood above is used to perform Bayesian inference on $\sigma$, which represents the structure function of the sample at a fixed time interval. The Hamiltonian Markov Chain Monte Carlo code Stan\footnote{https://mc-stan.org} is used to sample the likelihood of Equation \ref{eq:likelihood} and produce parameter posterior distributions. The free parameters of the model are the $\sigma$ of the normal distribution in Equation \ref{eq:likelihood} and the first epoch logarithmic flux  $\log f_{i,1}$ of individual pairs. During the sampling process we adopt a flat prior on $\sigma$ between $10^{-5}$ and $2$. The  $\log f_{i,1}$ is assigned a broad normal prior  with mean $-14$ (i.e. flux of $10^{-14}\rm \, erg \, s^{-1} \, cm^{-2}$) and (logarithmic) scatter parameter of 10.

Appendix \ref{sec:SF-validation} tests and validates the Bayesian SF estimation method described above using simulated light curves \citep{Sartori2018, Sartori2019} and X-ray photometric products that mimic those of the {\it XMM-Newton} and eRASS1-DE DR1 observations of SDSS DRQ16 QSOs.

\section{Results}\label{sec:results}

\subsection{The X-ray structure function of the full sample}

We first estimate the structure function of the full sample of DRQ16 QSOs described in Section \ref{sec:sample}. The results are plotted as a function rest-frame time scale, $\Delta T$,  in Figure \ref{fig:SF-FULL}. The SF monotonically increases with increasing $\rm \Delta T$ in agreement with previous studies \citep[e.g.][]{Vagnetti2016, Middei2017}. Assuming that the structure function is described by a power-law of the form $SF \propto \Delta T^\gamma$,  we estimate an index $\gamma=0.12\pm0.01$. This value translates to a Power Spectrum Density (PSD) power law index of about --1.2 \citep[e.g.][]{Bauer2009}. 

It is interesting to explore how these results compare with the predictions of empirical models for the PSD of AGN derived from detailed monitoring campaigns of local Seyferts. Because there is no analytic relation between PSD and SF, the generation of model predictions requires the intermediate step of generating simulated light curves for a given input PSD. The corresponding structure function at different rest-frame timescales can then be estimated numerically from these light curves. We use the method described in \cite{Emmanoulopoulos2013} as implemented in \cite{Sartori2018, Sartori2019} to generate light curve simulations. In this exercise we adopt a PSD model that follows a bending power-law  form \citep{McHardy2004, Gonzalez-Martin_Vaughan2012}

\begin{equation}\label{eq:psd}
PSD(\nu) = A\,\nu^{-1} \, \left(1+    \frac{\nu}{\nu_b}\right) ^{-1},
\end{equation}

\noindent where $\nu$ is the frequency. The amplitude, $A$, and break frequency, $\nu_b$, are suggested to depend on the physical parameters of the accreting system, e.g. the mass of the black hole and its Eddington ratio. Different observational studies of local AGN have proposed over the years different parameterisations for $A$ and $\nu_b$. In our analysis we use three popular models put forward in the literature. The first one, which we refer to as Model 1, assumes a constant amplitude $A = 2\cdot\nu_b \cdot PSD(\nu_b) = 0.02$ and break frequency, $v_b$, that depends on the black hole mass of the accreting system as


\begin{equation}\label{eq:model1-nub}
\nu_b = \frac{580}{M_{\rm BH}/M_\odot} \; (\rm s^{-1}).
\end{equation}

\noindent  This parametrisation is motivated by the observations of \cite{Papadakis2004} and  \cite{Gonzalez-Martin_Vaughan2012}.  Simulated light curves are generated \citep{Emmanoulopoulos2013, Sartori2018, Sartori2019} for two black hole masses $\log M_{BH}/M_{\odot} = 8, \; 9$ that bracket the mode of the distribution shown in Figure \ref{fig:MBHLEDD-dist}. The structure function that is numerically determined from these light curves is overplotted in Figure \ref{fig:SF-FULL}. It is clear that the PSD Model 1 predictions are inconsistent with the measured  structure function of SDSS DR16 QSOs. Modifications to this model are therefore required to account for this inconsistency. One possibility is to consider different dependences of the PSD break frequency  and and amplitude on black hole mass and Eddington ratio. We test two variants of Model 1. The first adopts the constant PSD amplitude of Model 1 and further assumes that the break frequency is a function of both the black-hole mass and the accretion rate as proposed by \cite{McHardy2006}. This dependence  is expressed in terms of the AGN bolometric luminosity ($L_{\rm bol}$ ) as


\begin{equation}\label{eq:model2-nub}
\nu_b=\frac{200}{86400} \cdot \frac{L_{\rm bol}}{10^{44} {\rm erg\,s^{-1}}}\cdot \left(\frac{M_{\rm BH}}{10^6\,M_\odot}\right)^{-2}  \; (\rm s^{-1}).
\end{equation}

\noindent We refer to this PSD parametrisation as Model 2. We also consider a parameterisation in which the assumption of a constant PSD amplitude is relaxed. This is motivated by observations of local Seyferts suggesting that the amplitude of the PSD scales with Eddington ratio, $\lambda_{Edd}$, as \citep{Ponti2012}

\begin{equation}\label{eq:model3-amp}
A = 2\cdot\nu_b \cdot PSD(\nu_b) = 3\times10^{-2}\cdot\lambda_{Edd}^{-0.8}.
\end{equation}

\noindent In this particular parameterisation, which is referred to as Model 3, the break frequency depends on black hole mass as in Equation \ref{eq:model1-nub}. Figure \ref{fig:SF-FULL} shows the structure function of Models 2 and 3 for black hole masses $\log M_{BH}/M_{\odot} = 8, \; 9$ and Eddington ratio $\lambda_{Edd}=0.1$ that approximately corresponds to the mode of the distribution in Figure \ref{fig:MBHLEDD-dist}. For this choice of physical parameters the Model 2, 3 structure function predictions are in better agreement with the observations, particularly for $\log M_{BH}/M_{\odot} = 8$.  We emphasise that the model curves plotted in Figure \ref{fig:SF-FULL} are not fits to the structure function data points. Instead they help visualise the expectations of empirical models derived from extensive observational studies of nearby active black holes. In any case, the ensemble X-ray variability properties of optically selected QSOs at cosmological distances provide strong evidence that the PSD of AGN depends on both black hole mass and Eddington ratio. It is therefore interesting to further explore such a dependence by directly measuring the structure function of DRQ16 QSOs grouped by black hole mass and Eddington ratio.

\subsection{Black hole mass dependence of the X-ray structure function}

We select DRQ16 QSOs in a narrow Eddington ratio interval, $-1.25<\log\lambda_{Edd}<-0.5$ that brackets the mode of the distribution in Figure \ref{fig:MBHLEDD-dist}. This sub-sample allows us to explore variations of the structure function with black hole mass at nearly fixed Eddington ratio.  The choice of the Eddington ratio interval above is a trade-off between sufficiently large sample size and the desire to minimize the impact of Eddington ratio on the measured structure function. We further split this sub-sample into two nearly equal size black hole mass bins at the threshold $\log M_{BH}/M_{\odot} = 8.75$. The structure function is estimated independently for the two sub-samples and the results are shown in Figure \ref{fig:SF-BH}. DRQ16 QSOs with lower black-hole masses have a systematically higher structure function compared to more massive ones. This difference is not sensitive to the choice of black hole mass bins. Splitting for example the sample into three nearly equal size groups of black hole mass yields similar results, i.e. the structure function is systematically lower for QSOs with the most massive black holes (see Fig. \ref{fig:SF-MBH-WIDE} and discussion in Appendix \ref{ap:sf_wide}). Moreover, the observed trend is not dominated by a few high signal-to-noise ratio  sources in the sample but is instead representative of the full population. We test this by re-running the analysis after removing observations in which individual QSOs have more than 100 X-ray counts (36\% of pairs removed) or 250 X-ray counts (12\% of the pairs removed). It is found that the main trend shown in Figure \ref{fig:SF-BH}  holds for these smaller sub-samples, i.e. a systematically higher SF for lower black hole masses. 

Although the structure functions of both sub-samples plotted in Figure \ref{fig:SF-BH}  increase on average with increasing time-scale, there is scatter among neighboring data points that is larger than their errorbars. Moreover, there is one  time-scale bin in Figure \ref{fig:SF-BH} for which the general trend of the SF with $M_{BH}$ is inverted, i.e. the high black hole mass sub-sample is more variable. Uncertainties in the determination of black-hole masses \cite[$\approx0.5$\,dex,][]{Wu_Shen2022} may weaken or even invert any intrinsic correlation between structure function amplitude and black hole mass. 

Additionally, although we attempt to control for variations of parameters that are important for the determination of the SF, e.g. black-hole mass, Eddington ratio or rest-frame time-scale,  by selecting QSOs in narrow parameter slices, the distribution of these parameters within individual time-scale and black-hole bins may be different, thereby leading to increased scatter. We demonstrate this point in Figure \ref{fig:diagnostic-bh} that shows for each time-scale bin of the data points in Figure \ref{fig:SF-BH} the distribution of key parameters, such as $\Delta T$, $M_{BH}$, $\lambda_{Edd}$ and redshift for the two QSO sub-samples selected by black hole mass. It is clear from this figure that e.g. the mean  $\lambda_{Edd}$ of the two sub-samples is similar across different time-scale bins, but the underlying distributions span a range of values and in some cases show subtle differences.  This is further demonstrated by applying the K-S (Kolmogorov-Smirnov) test to compare the histograms shown in each panel of Figure \ref{fig:diagnostic-bh} and estimate the probability of the null hypothesis that they are drawn from the same parent population. For certain $\lambda_{Edd}$ panels the null hypothesis can be rejected with a probability of about $3\times10^{-3}$, i.e at approximately the $3\sigma$ confidence level. We caution against false positives in the case of hypothesis testing with multiple comparisons and the need to consider corrections to the estimated $p$-values to account for this effect (e.g. Bonferroni adjustment).

The K-S test shows that the lowest null hypothesis probabilities occur for the redshift panels of Figure \ref{fig:diagnostic-bh}, with the high black-hole mass sub-samples being typically skewed to higher redshifts as a result of observational selection effects (e.g. flux limit, larger cosmological volumes). Variations of the mean redshift of a given black-hole mass subsample across different rest-frame time-scale bins are also visible in Figure \ref{fig:diagnostic-bh}. These effects translate to different rest-frame energies at which the variability of individual sources is estimated. Although the dependence of the variability amplitude on rest-frame energy is second order \cite[e.g.][]{Vagnetti2016, Ponti2012}, it may contribute to the scatter among different data points in Figure \ref{fig:SF-BH}. 

Also plotted in Figure \ref{fig:SF-BH} are the structure function predictions of PSD Models 2 and 3 for different black holes masses and Eddington ratios that correspond to the averages of the two observational subsamples. We re-iterate that these curves are not fits to the data but instead serve to guide the reader on the expectations based on our understanding of the variability of AGN from monitoring campaigns of local samples. Qualitatively these models are consistent with the observations, i.e. the corresponding structure functions at fixed timescale decrease with increasing black hole mass. Quantitatively however, the model structure functions are systematically lower than the observations. This suggests that small modifications may be necessary to the scaling factors that govern the relation between PSD amplitude and/or break frequency with black hole mass or Eddington ratio, i.e. Equations \ref{eq:model1-nub}, \ref{eq:model2-nub}, \ref{eq:model3-amp}.

\subsection{Eddington ratio dependence of the X-ray structure function}

\begin{figure}
	\includegraphics[width=\columnwidth]{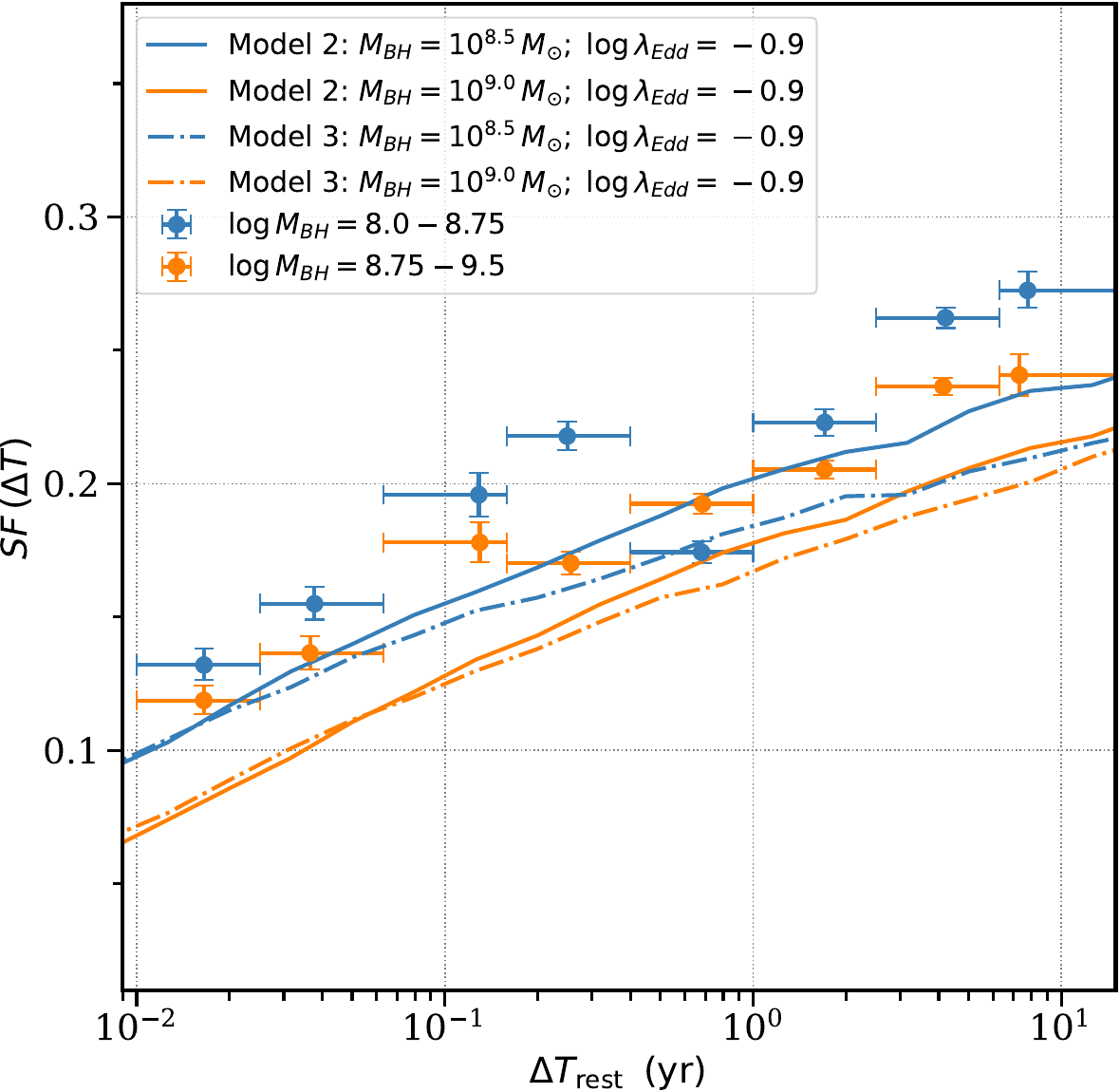}
    \caption{Structure function versus rest-frame time-scale for the SDSS DRQ16 QSO subsample with Eddington ratios in the interval $-1.25<\log\lambda_{Edd}<-0.5$, which is further split by black hole mass. The orange circles correspond to SDSS DRQ16 QSOs with $\log M_{BH} / M_\odot>8.75$. Blue circles are for QSOs with  black holes lower than the limit above, i.e. $\log M_{BH} / M_\odot<8.75$. The horizontal error bars show the extent of the rest-frame time interval. Data points are plotted at the mean $\Delta T_{rest}$ of a given sub-sample. Also shown are the SF predictions for the PSD Models 2 (solid lines; Eq. \ref{eq:psd}, \ref{eq:model2-nub}) and 3 (dashed lines; Eq. \ref{eq:psd}, \ref{eq:model1-nub}, \ref{eq:model3-amp}). The blue curves correspond to a black hole mass  $\log M_{BH}/M_{\odot} = 8.5$, while the orange curves assume $\log M_{BH}/M_{\odot} = 9$. The adopted Eddington ratio for all curves is $\log \lambda_{Edd} = -0.9$.}
    \label{fig:SF-BH}
\end{figure}

\begin{figure}
	\includegraphics[width=\columnwidth]{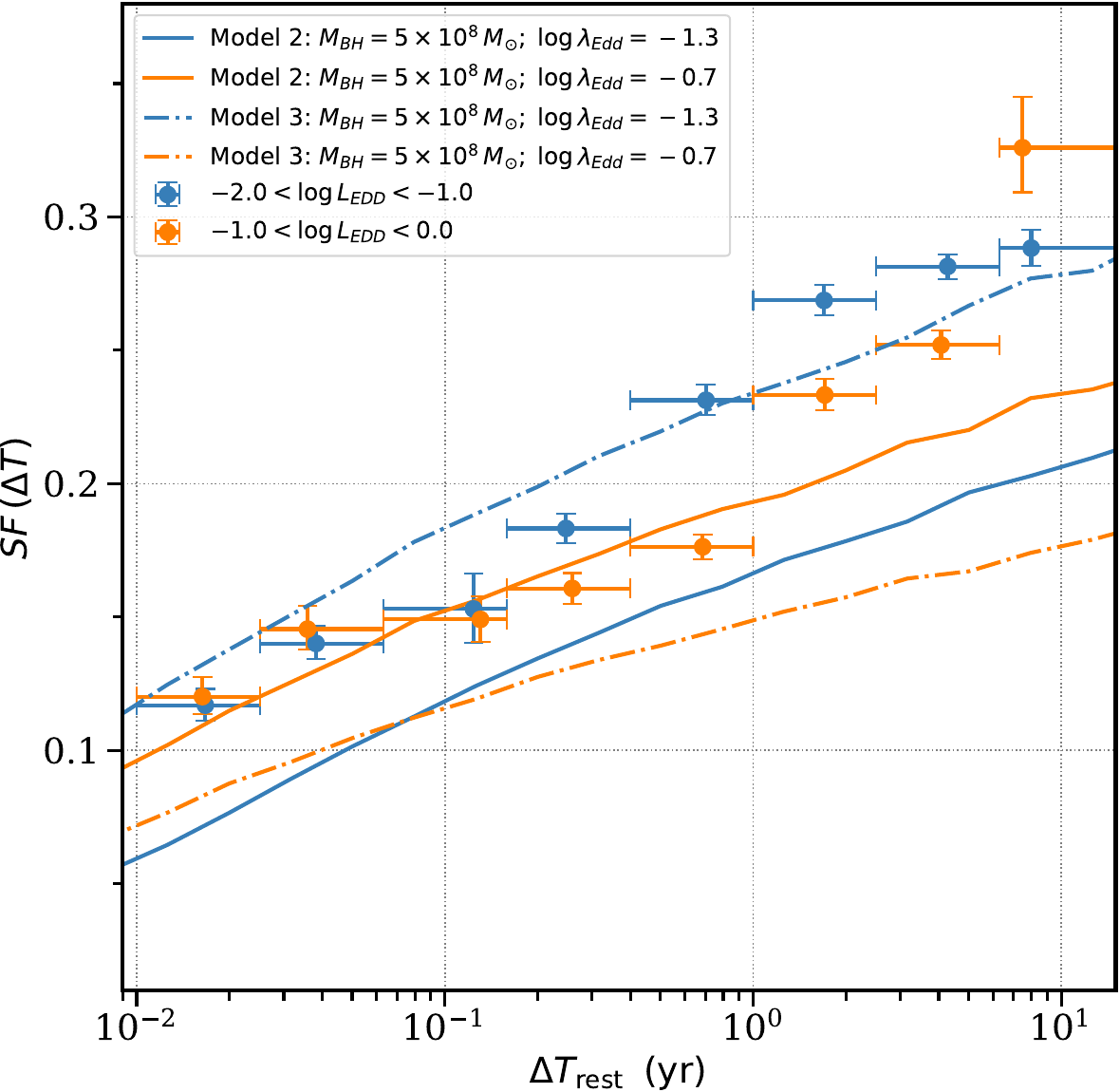}
    \caption{Structure function versus rest-frame time-scale for the subsample with $8.5<\log M_{BH}/M_{\odot}<9.0$, which is further split by Eddington ratio. The blue circles correspond to SDSS DRQ16 QSOs with $\log \lambda_{Edd} <-1$. Orange circles are for QSOs with  Eddington ratio larger than the limit above, i.e. $\log \lambda_{Edd} >-1$. The horizontal error bars show the extent of the rest-frame time interval. Data points are plotted at the mean $\Delta T_{rest}$ of a given sub-sample. Also shown are the SF predictions for the PSD Models 2 (solid lines; Eq. \ref{eq:psd}, \ref{eq:model2-nub}) and 3 (dashed lines; Eq. \ref{eq:psd}, \ref{eq:model1-nub}, \ref{eq:model3-amp}). The blue curves correspond to an Eddington ratio  $\log \lambda_{Edd} = -1.3$, while the orange curves assume $\log \lambda_{Edd} = -0.7$. The adopted black hole mass for all the curves is $\log M/M_{\odot} = 9$.}
    \label{fig:SF-LEDD}
\end{figure}

Next, we select DRQ16 QSOs in a narrow black-hole mass interval, $8.5<\log M_{BH}/M_{\odot}<9.0$ that brackets the mode of the distribution in Figure \ref{fig:MBHLEDD-dist}. This sub-sample is used to explore variations of the structure function with Eddington ratio at nearly fixed black hole mass. We define two Eddington ratio bins split at $\log \lambda_{Edd} = -1$. At this cut the two sub-samples have nearly equal number of 2-epoch X-ray observation pairs. The structure function is estimated for the two sub-samples and the results are plotted in Figure \ref{fig:SF-LEDD}. There are four time scale bins where the structure function of low Eddington ratio DRQ16 QSOs is higher compared to that of high Eddington ratio sources. For three time scale bins the structure functions of the two samples are consistent within the uncertainties. Only at the highest time-scales probed by the current sample ($\approx10$\,yrs) is the structure function of the high Eddington ratio sources larger than that of low Eddington ratio systems. Although the differences between the two sub-samples in Figure \ref{fig:SF-LEDD} are not as clear as in the case of black-holes mass bins (i.e. Figure \ref{fig:SF-BH}), there is evidence that low Eddington ratio DRQ16 QSOs tend to be more variable than high Eddington ratio systems, at least for time-scales between about 0.1--5\,yrs. Splitting the sample into three nearly equal size groups of Eddington ratio yields qualitatively similar conclusions, i.e. the structure function tends to be higher for DRQ16 QSOs with low Eddington ratios  (see Fig. \ref{fig:SF-LEDD-WIDE} and discussion in Appendix \ref{ap:sf_wide}). 
The same trend is also recovered after removing sources with a high number of counts, e.g. $>100$ (40\% of pairs removed). We reiterate that uncertainties in the estimation of black hole masses propagate into the Eddington ratio determination and could wash out intrinsic correlations between amplitude of variability and Eddington ratio. 

\begin{figure*}
	\includegraphics[width=1.99\columnwidth]{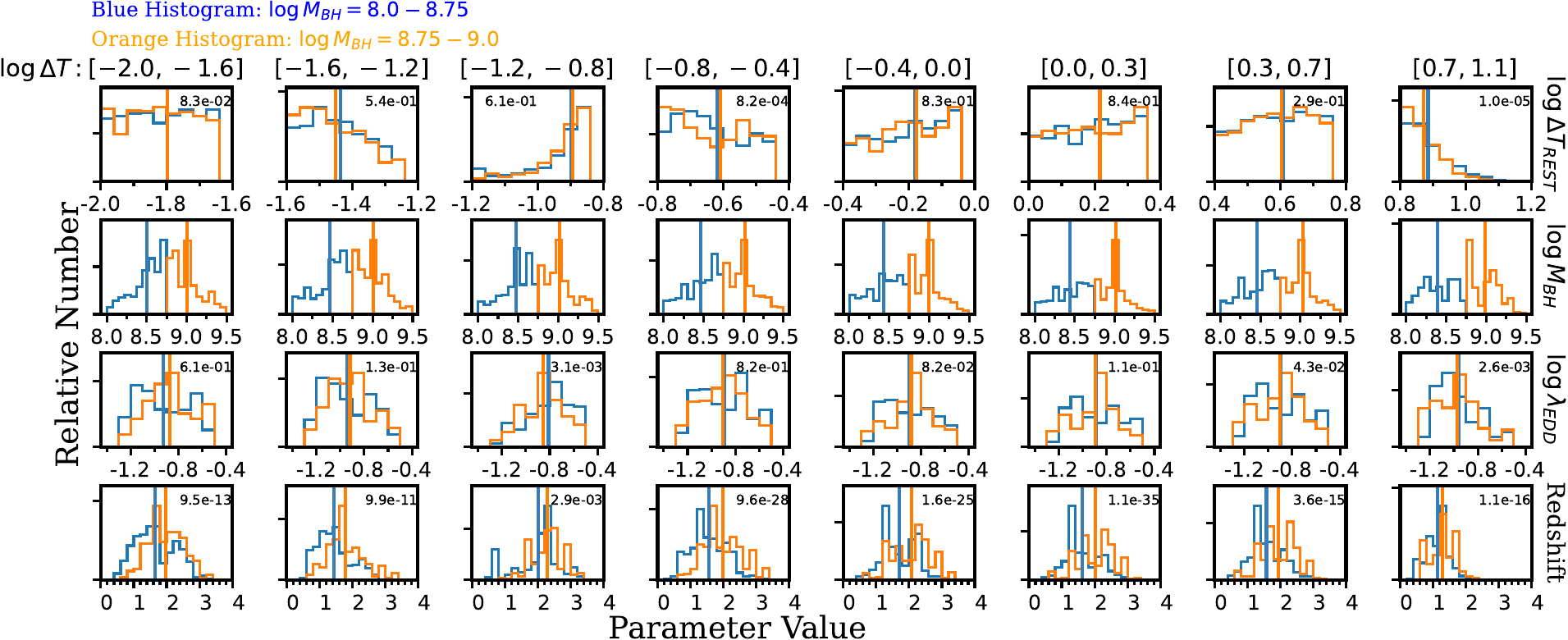} 
    \caption{The histograms in each panel show the distributions of each of the key  parameters, $\log \Delta T$ (top row), $\log M_{BH}$ (2nd row from top),  $\log \lambda_{Edd}$ (3rd row from top) and redshift (last row), for the two subsamples with Eddington ratio in the range $-1.25<\lambda_{Edd}<-0.5$ and black holes masses  in the intervals $\log M_{BH}/M_{\odot}= 8-8.75$ (blue lines) and $\log M_{BH}/M_{\odot}= 8.75-9.5$ (orange lines). Each column of panels corresponds to one  of the rest-frame times scales of the data points plotted in Fig. \ref{fig:SF-BH} as indicated at the top of the column. The vertical lines in each panel show the mean of the distribution of the corresponding parameter. The number in each panel corresponds to the Kolmogorow-Smirnow test p-value, i.e. the probability of the null hypothesis that the blue and orange histograms in each panel are drawn from the same parent distribution.}
\label{fig:diagnostic-bh}
\end{figure*}

\begin{figure*}
	\includegraphics[width=1.99\columnwidth]{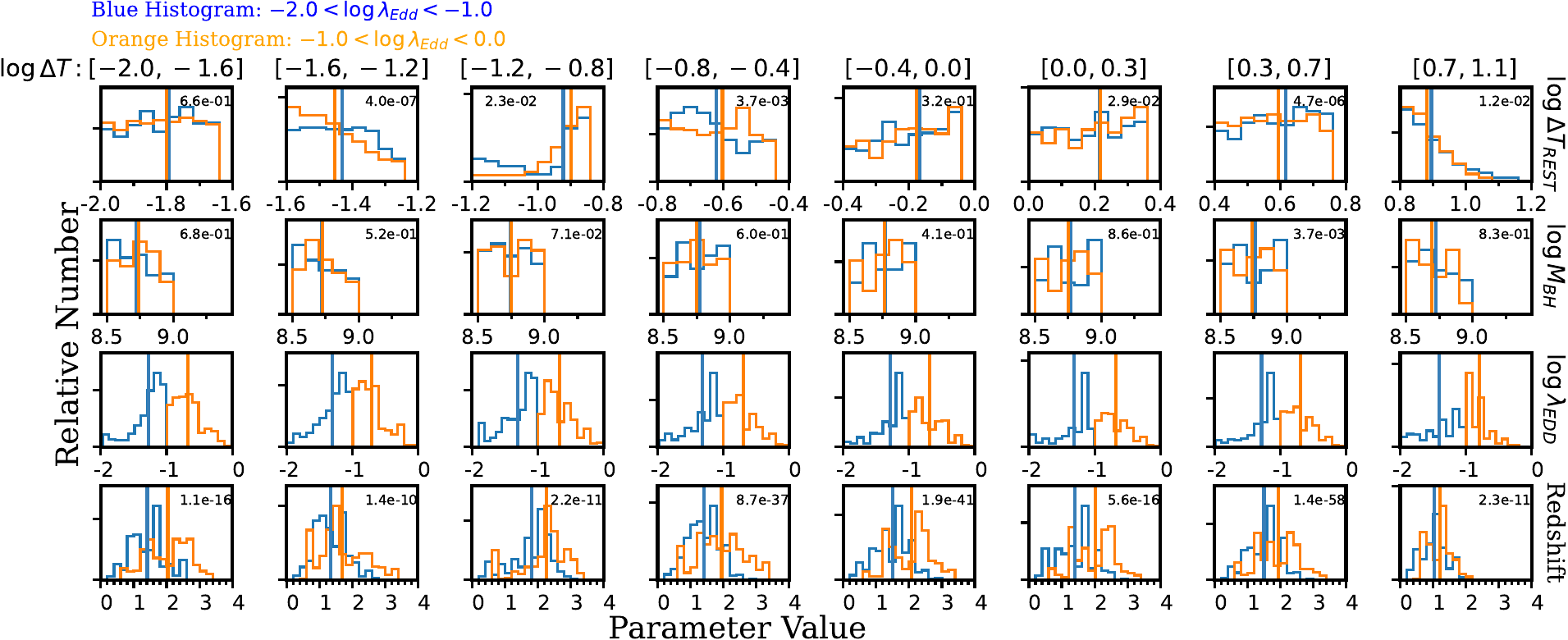}
    \caption{The histograms in each panel show the distributions of each of the key  parameters, $\log \Delta T$ (top row), $\log M_{BH}$ (2nd row from top),  $\log \lambda_{Edd}$ (3rd row from top) and redshift (last row), for the two subsamples with black holes masses  in the range $\log M_{BH}/M_{\odot}= 8.5-9.0$ and Eddington ratios in the intervals $-2.0<\log \lambda_{Edd}<-1.0$ (blue lines) and $-1.0<\log \lambda_{Edd}<0.0$ (orange lines). Each column of panels corresponds to one  of the rest-frame times scales of the data points plotted in Fig. \ref{fig:SF-LEDD} as indicated at the top of the column. The vertical lines in each panel show the mean of the distribution of the corresponding parameter. The number in each panel corresponds to the Kolmogorow-Smirnow test p-value, i.e. the probability of the null hypothesis that the blue and orange histograms in each panel are drawn from the same parent distribution.}
    \label{fig:diagnostic-ledd}
\end{figure*}

Figure \ref{fig:diagnostic-ledd} plots for each time-scale bin of Figure \ref{fig:SF-LEDD} and for each of the two QSO sub-samples selected by Eddington ratio the corresponding histograms of the quantities $\Delta T$, $M_{BH}$, $\lambda_{Edd}$ and redshift. We observe similar trends with black-hole mass, redshft and rest-frame time-scale as those already discussed in the case of Figure \ref{fig:diagnostic-bh}. Therefore part of the scatter observed in Figure \ref{fig:SF-BH} among neighboring data points is related to differences in the distribution of the parameters above for the various sub-samples. 

Also plotted in Figure \ref{fig:SF-LEDD} are the structure function predictions of PSD Models 2 and 3 for the mean Eddington ratios and black hole masses of the two observational subsamples. In Model 2 the structure function decreases with decreasing Eddington ratio, which is at odds with the observational results in Figure \ref{fig:SF-LEDD}. Model 3 instead, is qualitatively consistent with the observations predicting a higher structure function amplitude for lower Eddington ratios. Quantitatively, although this model is in reasonable agreement with the structure function measurements of the low Eddington-ratio sub-sample (mean $\log \lambda_{Edd}=-1.3$), it underestimates the structure function of the high Eddington-ratio sub-sample (mean $\log \lambda_{Edd}=-0.7$). This suggests that modifications need to be applied to Model 3 to make it quantitatively consistent with the measurements in Figure \ref{fig:SF-LEDD}. Exploring such changes is beyond the scope of this paper.


\begin{figure}
	\includegraphics[width=\columnwidth]{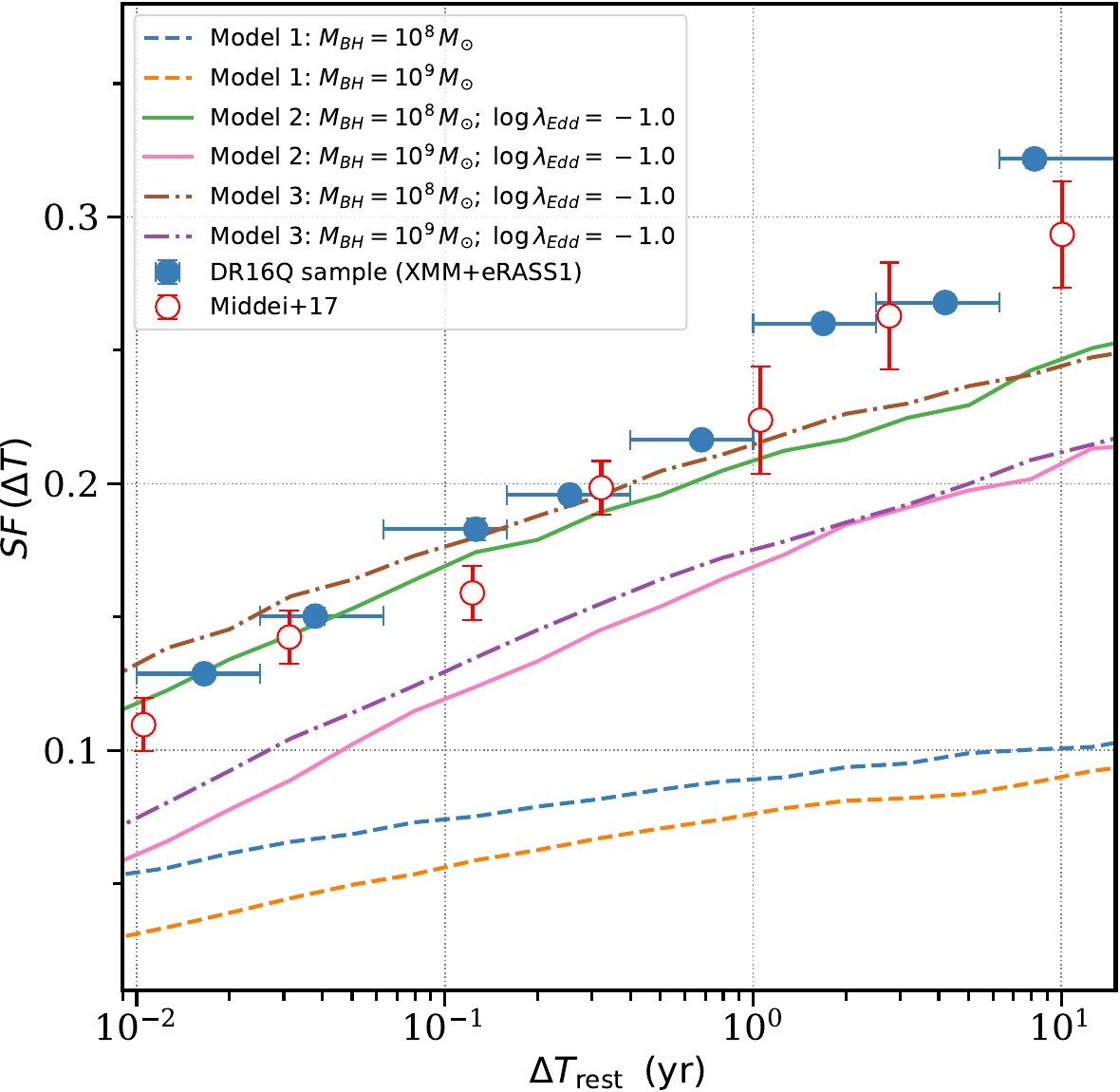}
    \caption{Structure function versus rest-frame time-scale. The blue circles are the results for SDSS DRQ16 QSOs based on the combined {\it XMM-Newton} and eRASS1-DE DR1 observations. The horizontal error bars show the extent of the rest-frame time interval. Data points are plotted at the mean $\Delta T_{rest}$ of a given  sub-sample. The observations are compared against the SF estimates presented by \protect\cite{Middei2017} for SDSS DRQ12 QSOs \citep{Paris2017} using {\it XMM-Newton} and ROSAT observations. Also shown are the structure function predictions of the PSD models 1, 2 and 3 parametrised by Equations \ref{eq:psd}-\ref{eq:model3-amp}.}
    \label{fig:SF-FULL}
\end{figure}

\section{Discussion}

We present a new Bayesian methodology for measuring the X-ray structure function of large AGN samples and apply it to SDSS DRQ16 QSOs with repeat observations from {\it XMM-Newton} and/or eROSITA. This dataset enables measurement of the X-ray structure function to decadal timescales as well as investigations on the dependence of the AGN long-term flux variability on black hole mass and Eddington ratio. 

An interesting result from our analysis is that we do not find evidence for a turnover in the structure function of SDSS DR1Q16 QSOs out to the longest time-scales ($\approx10$\,yr rest-frame) probed by the current sample. Instead Figure \ref{fig:SF-FULL} shows a monotonic increase of the structure function toward longer time intervals with a (logarithmic) slope of about 0.12. This is consistent with the results of \cite{Middei2017} on the X-ray structure function of SDSS DRQ12 QSOs based on combined ROSAT and {\it XMM-Newton} observations \citep[see also][]{Vagnetti2016}. Studies of the variability properties of QSOs in the UV/optical part of the electromagnetic spectrum also suggest a monotonically increasing structure function to decadal rest-frame time intervals with logarithmic slopes similar to those estimated by our analysis \citep[][but see \citealt{MacLeod2012A} for possible evidence for flattening]{deVries2005}. These findings have implications on the size of accretion disk and the origin of the long-term variability at X-ray wavelengths. 
In particular, there is an increasing body of evidence suggesting that the UV/optical variability of QSOs has two components that dominate at different temporal scales \citep[e.g.][]{Czerny2006, Breedt2009, HernandezSantisteban2020, Yao2023, Secunda2023}. At short time intervals the reprocessing of the stochastic X-ray flux variations associated with changes in the X-ray emitting corona are believed to drive the UV/optical variability \citep[e.g.][]{Kammoun2023}. On longer time-scales magneto-rotational turbulence \citep{BalbusHawley1991} or turbulence generated by convection as a result of opacity variations of the disk material \citep{JiangBlaes2020} introduce accretion disk instabilities that are thought to propagate inwards on the viscous timescale thereby introducing an additional long-term modulation of the light curves of AGN. Such long term variations may also affect the observed X-ray flux variability \citep[e.g.][]{PapadakisNandraKazanas2001, ArevaloUttley2006, Yao2023, Secunda2023} and lead to the monotonically increasing structure function of Figure \ref{fig:SF-FULL}.

Next we turn to the dependence of the stochastic flux variations of AGN on the physical parameters of the accreting system such as black hole mass and Eddington ratio. Such dependences provide a handle on the physical processes that drive the observed variability \cite[e.g.][]{Tang2023} and could enable the unification of active black hole systems with diverse masses and Eddington ratios under the same theoretical paradigm \cite[e.g.][]{McHardy2006}. Figures \ref{fig:SF-BH} and \ref{fig:SF-LEDD} indicate that the amplitude of the X-ray structure function of SDSS DRQ16 QSOs decreases with increasing black hole mass and toward higher Eddington ratios. This is consistent with the recent X-ray structure function calculations presented by \cite{Prokhorenko2024} based on SDSS DRQ14 QSOs \citep{Paris2018} with {\it XMM-Newton}/eROSITA overlapping observations in the part of the sky with proprietary rights belonging to the Russian SRG/eROSITA team. Our results are also consistent with  results on the dependence of the UV/optical variability of SDSS QSOs on the physical properties of the accreting system \citep[e.g.][]{Arevalo2023}. Similar trends, but on order of magnitude  shorter timescales (typically fraction of a day), are also recovered for the X-ray variability of local Seyferts based on either the normalised excess variance \citep{Ponti2012, Akylas2022} or the Power Spectral Density of the observed variations \citep{McHardy2004, McHardy2006}. 

It is also interesting to explore different empirical scaling relations that can potentially describe the observed dependence of the structure function on physical parameters. The fundamental quantity in this exercise is the PSD of the observed variations that describes the variability amplitude as a function of Fourier frequency. We first assume that the PSD is described by the bending power-law  of Eq. \ref{eq:psd} with parameters (i.e. break frequency, amplitude) that scale with black hole mass and/or Eddington ratio. We then consider different parametrisations for these relations derived from detailed studies of the temporal properties of local Seyeferts. It is shown that PSD models with constant amplitude and a break frequency that is only a function of black hole mass \citep[see Eq. \ref{eq:model1-nub},][]{Papadakis2002, UttleyMcHardyPapadakis2002} are inconsistent with the X-ray variability of distant QSOs (see Fig. \ref{fig:SF-FULL}). Moreover, PSD models in which $\nu_b$ scales with both black hole mass and Eddington ratio \citep[see Eq. \ref{eq:model2-nub},][]{McHardy2004, McHardy2006} predict a dependence of the structure function on Eddington ratio that is at odds with the observations in Fig. \ref{fig:SF-LEDD}. The PSD model that better describes the X-ray temporal properties of DRQ16 QSOs, at least qualitatively, is the one proposed by \cite{Ponti2012} with an amplitude that is inversely proportional to Eddington ratio (Eq. \ref{eq:model3-amp}) and a break frequency that decreases with increasing black hole mass as in Eq. \ref{eq:model1-nub}. 

Similar conclusions are reached from studies of the ensemble variability of a different sample of high redshift active supermassive black holes, X-ray selected AGN in the Chandra Deep Field South \citep{Luo2017}. The mean normalised excess variance of these sources is found to decrease with increasing X-ray luminosity  \citep{Paolillo2017}. Modelling suggests that this dependence cannot be reproduced by a PSD model with a constant amplitude and a break frequency given by either Equation \ref{eq:model1-nub} or Equation \ref{eq:model2-nub} \citep{Georgakakis2021}. Instead, a PSD amplitude that scales with Eddington ratio as in Equation \ref{eq:model3-amp} provides a better description of the ensemble variability of X-ray selected AGN in the Chandra Deep Field South \citep{Georgakakis2021}. 
Recently, \cite{Paolillo2023} combined archival observations of local Seyferts and higher redshift X-ray selected broad optical emission-line AGN detected in deep extragalactic surveys (COSMOS, Chandra Deep Field South) to measure the ensemble normalised excess variance of the population at different time scales and as a function of black hole mass. They adopt a PSD model with a constant amplitude (i.e. independent of redshift or Eddington ratio) and a break frequency that scales with  black hole mass to demonstrate that it provides a satisfactory description of their variability data. Nevertheless, \cite{Paolillo2023} do not explore the consistency of their observations with more complex PSD model parameterisations that e.g. include a dependency of the amplitude on Eddington ratio. Further work is therefore needed to compare our results and conclusions with those of \cite{Paolillo2023}. At any rate, our analysis demonstrates the power of ensemble X-ray flux variability studies of high redshift AGN samples as a means of constraining empirical PSD models. 

Dedicated spectroscopic follow-up programmes of the eROSITA X-ray sources, e.g. SDSS-V Black Hole Mapper \citep{Kollmeier2017, Almeida2023}, 4MOST AGN Survey \citep{Merloni2019}, in combination with the multi-epoch X-ray photometry from the successive eROSITA All Sky Surveys (eRASS1--5), will significantly increase the sample of AGN for ensemble variability investigations. These datasets can help refine and improve the observational trends presented in this work by both increasing the statistics and providing a wider black-hole mass and Eddington ratio baselines. 

\section{Conclusions}

This paper revisits the X-ray structure function of SDSS QSO samples using a novel Bayesian methodology and combining {\it XMM-Newton} archival data with new eROSITA observations carried out in the first six months of the SRG/eROSITA all-sky survey (eRASS1) and whose proprietary rights lie with the German eROSITA consortium. 

We find no evidence for a turnover in the X-ray structure function to the longest time-scales of about 10\,yr rest-frame probed by our sample. This finding may have implications on the size of the accretion disk, under the assumption  that the long-term variations of the accretion flow drive the variability at X-ray wavelengths on decadal timescales.

There is also evidence that the structure function of DRQ16 QSOs on timescales of few months to years increases with decreasing black hole mass and toward lower Eddington ratios. This dependence is best represented by PSD models with an amplitude that scales with Eddington ratio. 

Our analysis highlights the importance of ensemble variability studies as a complement to the characterisation of the temporal properties of individual AGN. 

\section*{Acknowledgements}
The authors thank the anonymous referee for their careful reading of the paper and their constructive comments. The research leading to these results has received funding from the Hellenic Foundation for Research and Innovation (HFRI) project "4MOVE-U" grant agreement 2688, which is part of the programme "2nd Call for HFRI Research Projects to support Faculty Members and Researchers". This work is based on data from eROSITA, the soft X-ray instrument aboard SRG, a joint Russian-German science mission supported by the Russian Space Agency (Roskosmos), in the interests of the Russian Academy of Sciences represented by its Space Research Institute (IKI), and the Deutsches Zentrum für Luft- und Raumfahrt (DLR). The SRG spacecraft was built by Lavochkin Association (NPOL) and its subcontractors, and is operated by NPOL with support from the Max Planck Institute for Extraterrestrial Physics (MPE). The development and construction of the eROSITA X-ray instrument was led by MPE, with contributions from the Dr. Karl Remeis Observatory Bamberg and ECAP (FAU Erlangen-Nuernberg), the University of Hamburg Observatory, the Leibniz Institute for Astrophysics Potsdam (AIP), and the Institute for Astronomy and Astrophysics of the University of Tübingen, with the support of DLR and the Max Planck Society. The Argelander Institute for Astronomy of the University of Bonn and the Ludwig Maximilians Universität Munich also participated in the science preparation for eROSITA. This research made use of  Astropy,\footnote{\href{www.astropy.org}{www.astropy.org}} a community-developed core Python package for Astronomy \citep{astropy:2013, astropy:2018}. Based on observations obtained with XMM-Newton, an ESA science mission with instruments and contributions directly funded by ESA Member States and NASA.

\section*{Data Availability}

The code and data used in this paper are available at GitHub \url{https://github.com/ageorgakakis/StructureFunction} and Zenodo \url{https://zenodo.org/record/10560969}.



\bibliographystyle{mnras}
\bibliography{mybib} 




\appendix

\section{Validation of the Bayesian structure function estimation method}\label{sec:SF-validation} 

We use simulated light curves with known statistical properties, i.e. power spectrum shape and normalisation, to validate the Bayesian approach of Section \ref{sec:SF} for the estimation of the structure function of QSO populations. 

The starting point of the validation exercise is a light curve generated by the code presented by \cite{Sartori2018, Sartori2019} based on the methods of \cite{Emmanoulopoulos2013}. For a given rest-frame time-scale, $\Delta T$, we can then randomly choose $N_{\rm sample}$ number of data point pairs on the light curve each of which are separated by the time interval $\Delta T$. These pairs represent the 2-epoch X-ray photometry of a  population of QSOs with size $N_{\rm sample}$ that can be used to determine their structure function. 

Before drawing  $N_{\rm sample}$ pairs however, mean fluxes need to be assigned to the light curves that are representative of the SDSS DRQ16 QSO sample.  Using the RapidXMM aperture photometry of the SDSS QSO sample we estimate 0.2-2\,keV fluxes for individual sources following the approach explained in Appendix \ref{ap:flux} and then build the distribution of the sample shown in Figure \ref{fig:flux-dist}. We use this historgam to randomly draw $N_{\rm sample}$ fluxes that are representative of the SDSS DRQ16 QSO sample. These are used to normalise the mean of the simulated light curves. The end product of this step are $N_{\rm sample}$ light curves with the same power spectrum but different mean fluxes. It is these $N_{\rm sample}$ renormalised light curves that we use to draw (from each) two random fluxes separated by the time interval $\Delta T$. These flux pairs are then converted to {\it XMM-Newton} EPIC (first epoch) and eROSITA (2nd epoch) photon counts or just {\it XMM-Newton} photon counts (both epochs). This conversion uses exposure times and backgrounds ({\it XMM-Newton} or eROSITA) randomly drawn from the RapidXMM or \texttt{apetool} X-ray aperture photometry products extracted at the positions of SDSS QSO as described in Section \ref{sec:sample}. This process yields simulated data sets that are similar to real observations that can be analysed using the Bayesian methodology of section \ref{sec:SF} to infer the structure function of the population at the time-scale $\Delta T$. The inferred SF can then be compared with the known structure function of the input light curve.

\begin{figure}
	\includegraphics[width=\columnwidth]{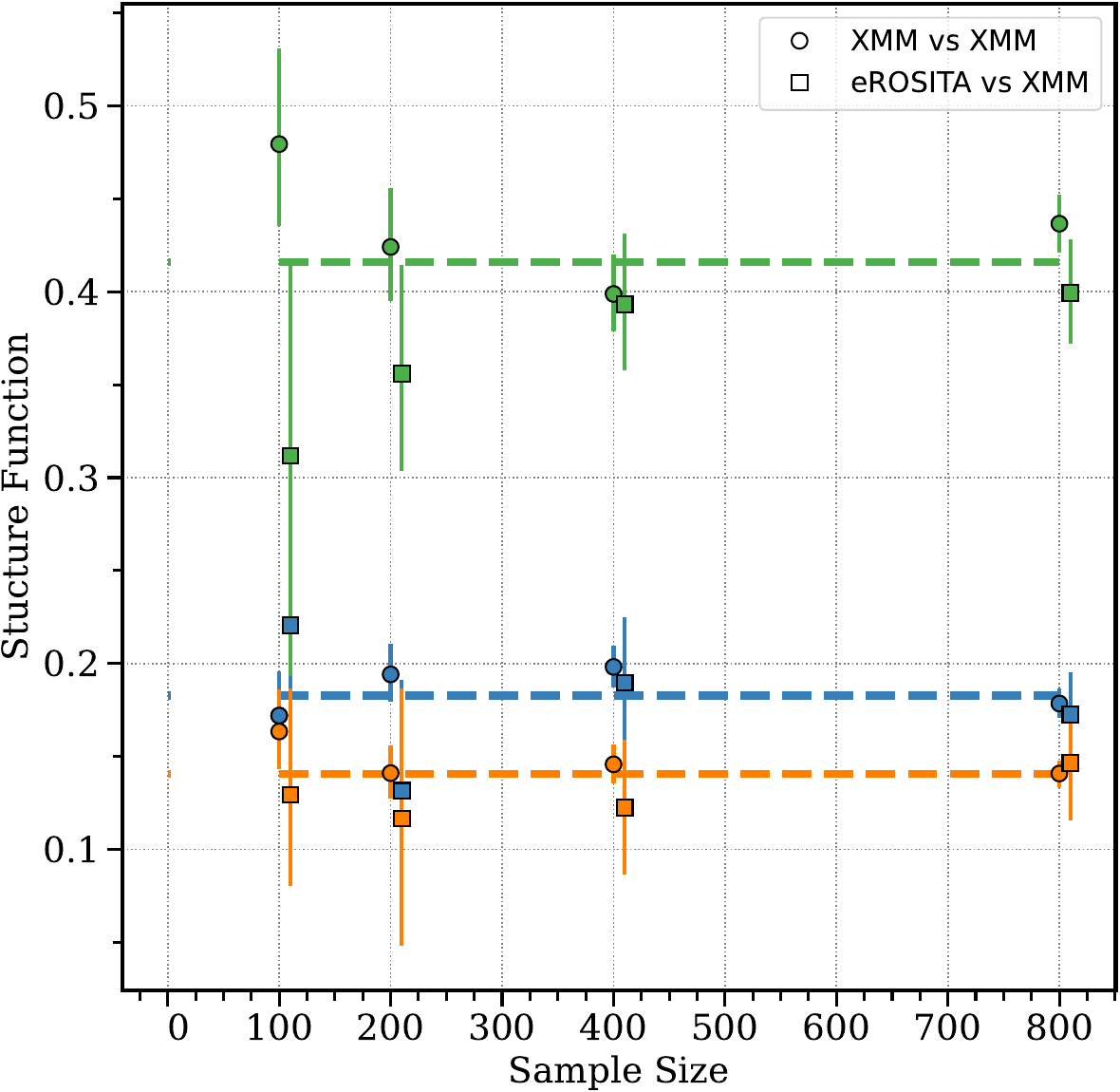}
    \caption{Inferred structure function from the simulations described in Appendix \ref{sec:SF-validation} as a function of $N_{sample}$, i.e. the simulated number of 2-epoch X-ray observation pairs. The squares are the inferred structure functions for simulations where the photon counts are from {\it XMM-Newton} EPIC for one of two epochs and eROSITA-DE DR1 for the second epoch. The circles show the inferred structure functions in the case of simulations in which  both epochs are assumed to be observed by {\it XMM-Newton} EPIC. The uncertainties associated with each data point correspond to $1\sigma$ and are determined from the posteriors of the Bayesian approach described in Section \ref{sec:SF}. The dashed lines correspond to 3 different input structure function levels that correspond to either different time intervals $\Delta T$ or different PSD normalisations. These input structure function levels are marked with different colors (orange, blue, green) and should be compared with the inferred structure function data points of the same colour. }
    \label{fig:sf-sim}
\end{figure}

Figure  \ref{fig:sf-sim} shows the results of this experiment by comparing the inferred SF with the input one as a function of the sample size, $N_{\rm sample}$. Two separate sets of data points are shown in this figure. The first combines {\it XMM-Newton} EPIC (first epoch) with  eROSITA-DE DR1 (2nd epoch) photon counts to infer the SF. The second assumes {\it XMM-Newton} photon counts for both epochs to derive the structure function. Moreover, different timescales $\Delta T$ and/or PSD normalisations are used as input, which result in three different values of the input structure function that are to be recovered. These are shown with the three horizontal lines in Figure  \ref{fig:sf-sim}. This figure demonstrates that when increasing $N_{\rm sample}$ the inferred SF converges to the input one. Additionally, the inferred SF is always consistent with the input one within the $1\sigma$ uncertainty estimated from the posteriors returned by the Bayesian methodology of Section \ref{sec:SF}. These results test and validate the Bayesian structure function calculation developed in Section \ref{sec:SF}.

\section{X-ray flux estimation using aperture photometry products}\label{ap:flux}

\begin{figure}
	\includegraphics[width=\columnwidth]{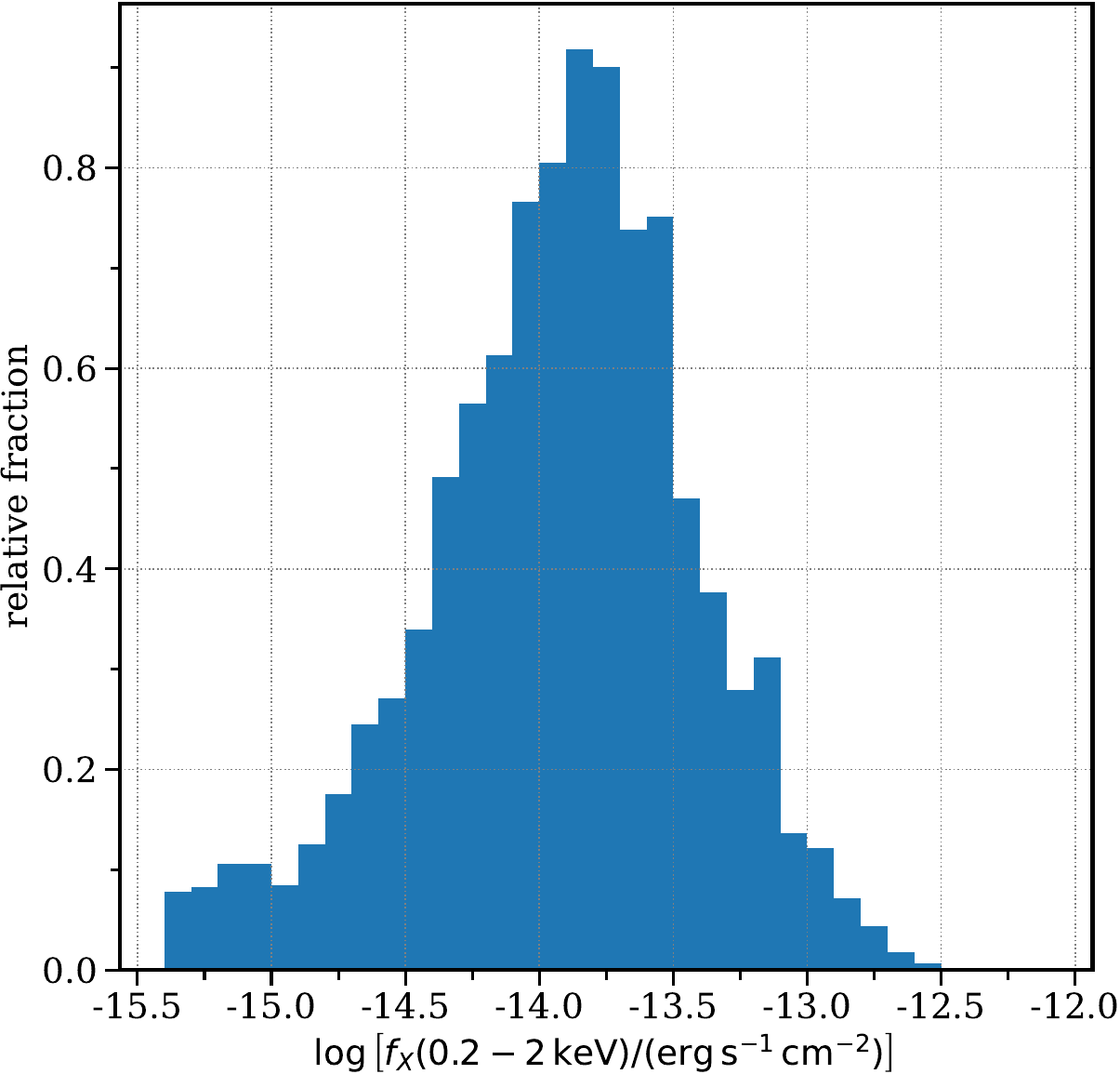}
    \caption{SDSS QSO 0.2-2\,keV X-ray flux distribution inferred from the RapidXMM aperture photometric products (see text for details).}
    \label{fig:flux-dist}
\end{figure}

This Section describes the determination of X-ray fluxes for the SDSS QSOs based on the RapidXMM aperture photometry. Suppose a source on the detector $i$ of a given {\it XMM-Newton} observation for which the RapidXMM database lists $N_i$ total number of photons, a background level $B_i$ and an exposure time $t_i$. The probability that this source has flux $f_X$ is given by the Poisson probability distribution function

\begin{equation}\label{eq:flux-poisson}
    P(f_{X}) = \frac{e^{-\lambda} \times\, \lambda^{N} } {N!},
\end{equation}

\noindent where $N=\sum_{i=1}^3 N_i$ is the sum of the source's photons in each of the three EPIC detectors.  The Poisson expectation value in the equation above can be written as

\begin{equation}\label{eq_sum_flux}
\lambda = f_X \times \sum_{i=1}^3 \bigl(  ECF_i  \times EEF_i \times t_i
\bigr) + \sum_i B_i,
\end{equation}

\noindent where $ECF_i$ is the count to flux conversion  factor for  the  camera $i$ and $EEF_i$  is the  encircled energy fraction of the RapidXMM aperture for the detector $i$ \citep[see Table 1 of][]{Ruiz2022}. The summation is over the three EPIC cameras, PN, MOS1 and MOS2.  The count to flux conversion factor is estimated for each EPIC camera separately  assuming  a power-law  X-ray  spectrum with  $\Gamma=1.9$ absorbed by the appropriate Galactic hydrogen column  density \citep{Kalberla2005}  in the direction of the SDSS QSO under consideration.  Equation \ref{eq:flux-poisson} can be solved numerically to determine the X-ray flux probability distribution function. This then used to determine the mode for each SDSS QSO and build the distribution plotted in Figure \ref{fig:flux-dist}. 

\section{Structure function for widely separated black hole and Eddington ratio bins}\label{ap:sf_wide}

In this section we explore the dependence of the structure function on black hole mass and Eddington ratio by adopting a different binning of the SDSS QSOs compared to Section \ref{sec:results}. Instead of adopting a single threshold that splits the sample into two groups of black hole mass or Eddington ratio (see Section \ref{sec:results}), we consider subsets that are widely separated in these two physical quantities. This is to minimise the impact of random and systematic uncertainties in the determination of black hole masses that add scatter and may result in the mixing up of QSOs around a given threshold. In practice we split the sample into 3 groups of black hole mass or Eddington and compare the structure function of the two non-adjacent subsets. 

\begin{figure}
	\includegraphics[width=\columnwidth]{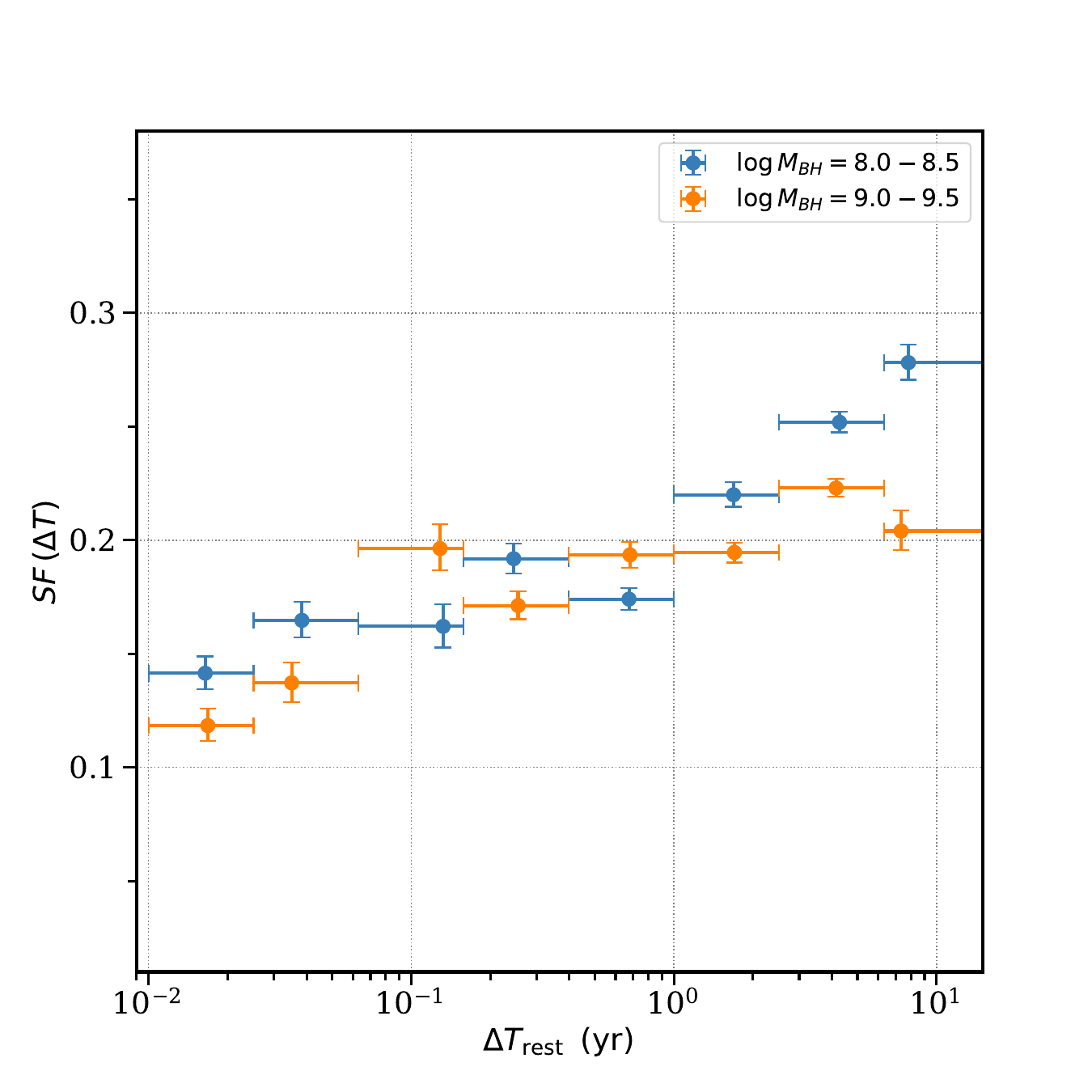}
    \caption{Structure function versus rest-frame time-scale for the SDSS QSO subsample with Eddington ratios in the range $-1.25<\log\lambda_{Edd}<-0.5$ and black hole masses in the intervals $\log M_{BH}/M_{\odot} = 8.0-8.5$ (blue data point) and $\log M_{BH}/M_{\odot}=9.0-9.5$ (orange data points). The vertical error bars correspond to the 68\% confidence interval uncertainties in the determination of the structure function. The horizontal error bars show the extent of the rest-frame time interval. Data points are plotted at the mean $\Delta T_{rest}$ of a given sub-sample.}
    \label{fig:SF-MBH-WIDE}
\end{figure}

First we select DRQ16 QSOs in the  Eddington ratio interval, $-1.25<\log\lambda_{Edd}<-0.5$, that brackets the mode of the distribution in Figure \ref{fig:MBHLEDD-dist}. We further split this sub-sample into three black hole mass bins, $\log M_{BH}/M_{\odot} = 8.0-8.5$, $8.5-9.0$ and $9.0-9.5$. Figure  \ref{fig:SF-MBH-WIDE} compares the structure function of the sub-smaples with  $\log M_{BH}/M_{\odot} = 8.0-8.5$ and $9.0-9.5$. Least massive black holes tend to have larger structure function amplitudes at fixed time scale, in agreement with the findings of 
 Figure \ref{fig:SF-BH}. Nevertheless, there are two time-scale bins in Figure  \ref{fig:SF-MBH-WIDE} for which this trend is inverted. 
 
\begin{figure}
	\includegraphics[width=\columnwidth]{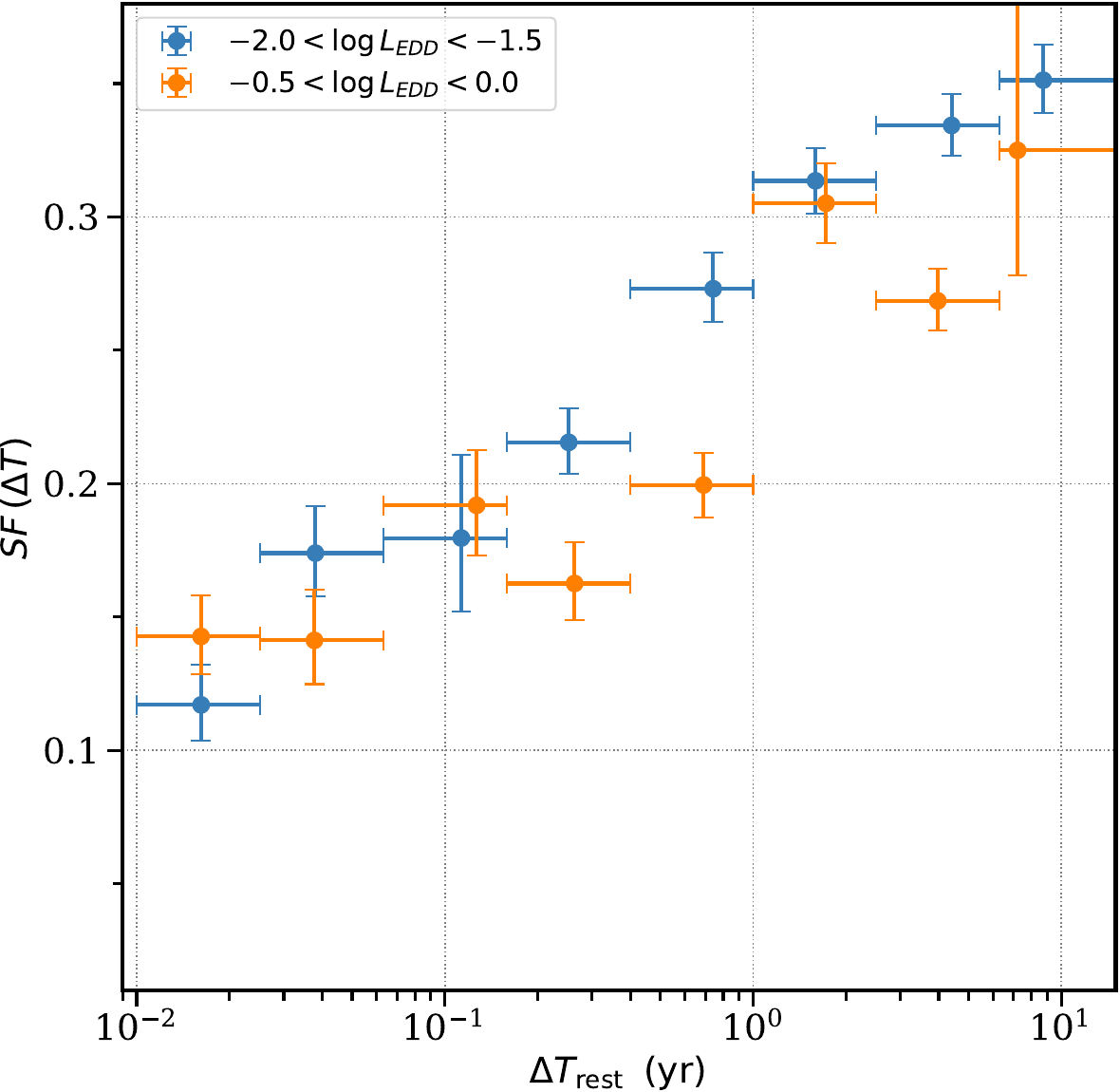}
    \caption{Structure function versus rest-frame time-scale for the SDSS QSO subsample with black hole masses in the range $8.5<\log M_{BH}/M_{\odot}<9.0$  and Eddington ratios in the intervals $-2<\log \lambda_{Edd}<-1.5$ (blue data points) and $-1<\log \lambda_{Edd}<0$ (orange data points). The vertical error bars correspond to the 68\% confidence interval uncertainties in the determination of the structure function. The horizontal error bars show the extent of the rest-frame time interval. Data points are plotted at the mean $\Delta T_{rest}$ of a given sub-sample.}
    \label{fig:SF-LEDD-WIDE}
\end{figure}

Next, we select DRQ16 QSOs in the narrow black-hole mass interval, $8.5<\log M_{BH}/M_{\odot}<9.0$ and further split them into three groups of Eddington ratio, $-2<\log \lambda_{Edd}<-1.5$, $-1.5<\log \lambda_{Edd}<-0.5$ and $-0.5<\log \lambda_{Edd}<0$. The results are plotted in Figure \ref{fig:SF-LEDD-WIDE} for the two non-adjacent bins, $-2<\log \lambda_{Edd}<-1.5$ and $-0.5<\log \lambda_{Edd}<0$. This figure supports the trend of a higher structure function toward lower Eddington ratios.

\bsp	
\label{lastpage}
\end{document}